\documentclass[letterpaper,twocolumn,10pt]{article}
\usepackage{usenix-2020-09}
\usepackage{enumitem}
\usepackage{xcolor}
\usepackage{graphicx}
\usepackage{tikz}
\usepackage{amsmath}
\usepackage{filecontents}
\usepackage{xspace}
\usepackage{amsmath}
\usepackage{color, soul}
\usepackage{multirow}
\usepackage{hyperref}
\usepackage{subfig}
\usepackage{booktabs}

\newcommand{\gc}[0]{GC\xspace}
\newcommand{\ser}[0]{S/D\xspace}
\newcommand{\name}[0]{\emph{TeraHeap}\xspace}
\newcommand{\mmio}[0]{mmio\xspace}

\newcommand{\comment}[1]{{}}
\newcommand*\circled[1]{\tikz[baseline=(char.base)]{
            \node[shape=circle,fill,inner sep=0.4pt] (char) {\textcolor{white}{#1}};}}

\author{
  {\rm Iacovos G. Kolokasis\textsuperscript{1}}\\
  Insitute of Computer Science, FORTH\\
  kolokasis@ics.forth.gr
  \and
  {\rm Giannos Evdorou\textsuperscript{1}}\\
  Insitute of Computer Science, FORTH\\
  evdorou@ics.forth.gr
  \and
  {\rm Anastasios Papagiannis}\\
  Insitute of Computer Science, FORTH\\
  apapag@ics.forth.gr
  \and
  {\rm Foivos Zakkak}\\
  Red Hat. Inc\\
  fzakkak@redhat.com
  \and
  {\rm Christos Kozanitis}\\
  Insitute of Computer Science, FORTH\\
  kozanitis@ics.forth.com
  \and
  {\rm Shoaib Akram}\\
  Australian National University\\
  shoaib.akram@anu.edu.au
  \and
  {\rm Polyvios Pratikakis\textsuperscript{1}}\\
  Insitute of Computer Science, FORTH\\
  polyvios@ics.forth.com
  \and
  {\rm Angelos Bilas\textsuperscript{1}}\\
  Insitute of Computer Science, FORTH\\
  bilas@ics.forth.gr
}

\begin{document}

%don't want date printed
\date{}

\title{Garbage Collection or Serialization? Between a Rock and a Hard Place!}

\maketitle

\footnotetext[1]{Also with the Computer Science Department, University of Crete}

\begin{abstract}

  Big data analytics frameworks, such as Spark and Giraph, need to
  process and cache massive amounts of data that do not always fit on
  the heap. Therefore, frameworks temporarily move long-lived objects
  outside the managed heap (off-heap) on a fast storage device.
  Unfortunately, this practice results in:
  (1) high serialization/deserialization (\ser) cost, and
  (2) high memory pressure when off-heap objects are moved
  back to the managed heap for processing.

  In this paper, we propose \name{}, a system that eliminates \ser
  overhead and expensive \gc scans for a large portion of the objects
  in big data frameworks. \name{} relies on three concepts.
  (1) It eliminates \ser cost by extending the managed runtime (JVM)
  to use a second high-capacity heap (H2) over a fast storage device.
  (2) It reduces \gc cost by fencing the garbage collector from
  scanning H2 objects.
  (3) It offers a simple hint-based interface, which allows
  frameworks to leverage knowledge about objects for populating H2.

  We implement \name{} in OpenJDK and evaluate it with 15 widely used
  applications in two real-world big data frameworks, Spark and
  Giraph. Our evaluation shows that for the same DRAM size, \name{}
  improves performance by up to 73\% and 28\% compared to native Spark
  and Giraph, respectively. Also, it provides better performance by
  consuming up to $8\times$ and $1.2\times$ less DRAM capacity than
  native Spark and Giraph, respectively. Finally, it outperforms
  Panthera, a garbage collector for hybrid memories, by up to 69\%.

\end{abstract}

\section{Introduction}

Managed big data frameworks, such as Spark~\cite{Zaharia:Spark} and
Giraph~\cite{giraph}, are designed to analyze huge volumes of data.
Typically, such processing requires iterative computations over data
until a convergence condition is satisfied. Each iteration produces
new transformations over data, generating a massive volume of objects spanning long
computations.

Hosting these objects on the managed heap increases memory pressure,
resulting in frequent garbage collection (GC) cycles with low yield.
Each \gc cycle reclaims little space because
(1) the cumulative volume of allocated objects is several times larger
than the size of available heap~\cite{Xu:Neutrino}, and
(2) objects in big data frameworks exhibit long
lifetimes~\cite{lifetime-ferreira,wang:Panthera,Xu:gc_evaluation}.
Although production garbage collectors efficiently manage short-lived
objects, they do not perform well under high memory pressure
introduced by long-lived objects~\cite{mitchel:bloat}.

Hence, the common practice of coping with the rapidly growing datasets
and high \gc cost is to move objects outside the managed heap
(off-heap) over a fast storage device (e.g., NVMe SSD). However,
frameworks cannot compute directly over off-heap objects, and thus,
they (re)allocate these objects on the managed heap to process them.
Although some systems support off-heap computation over byte arrays
with primitive types~\cite{arrow, spark_tungsten}, they do not offer
support for computation over arbitrary schema objects, which
applications use extensively.

Moving managed objects off-heap has two main limitations. First, it
introduces high serialization/deserialization (\ser) overhead for
applications that use complex data
structures~\cite{ibanez:zerializer,matei:champions,nguyen:skyway}.
Recent efforts~\cite{jang:arch,li:hods} reduce \ser but demand custom
hardware extensions and do not mitigate \gc overhead. Second, moving a
large volume of off-heap objects to the managed heap for processing
raises the \gc cost. Although TMO~\cite{weiner:tmo} transparently
swaps cold application memory to NVMe SSDs and provides direct access
to device resident objects (no \ser), it cannot avoid slow \gc scans
over the device.
Our evaluation shows that \gc and \ser constitute up to 87\%
of the execution time in big data applications.

In this work, we propose \name{}, a system that eliminates \ser and
\gc overheads for a large portion of the data in managed big data
analytics frameworks. \name{} extends the JVM to use a second,
high-capacity heap (H2) over a fast storage device that coexists
alongside the regular heap (H1). It eliminates \ser by providing
direct access to objects in H2 and reduces \gc by avoiding costly \gc
scans over objects in H2. Frameworks use \name{} through its
hint-based interface without modifications to the applications that
run on top of them. \name{} addresses three main challenges, as
follows.

\paragraph{Identifying candidate objects for H2:}
Big data frameworks move specific objects outside the managed heap on
off-heap storage. For instance, Spark moves off-heap cached
intermediate results; Giraph moves the graph's vertices, edges, and
messages. Frameworks organize such data (partitions) as groups of
objects with a single-entry root reference~\cite{mahmud:survey}.
\name{} provides a hint-based interface based on \emph{key-object
opportunism}~\cite{hayes:key_opport}, enabling frameworks to mark root
key-objects and indicate when to move them to H2. During \gc, \name{}
starts from root key-objects and dynamically computes the remaining
group objects for moving to H2.

\paragraph{Eliminating \gc cost for H2:}
\name{} presents a unified heap (H1+H2), where \emph{scans over H2
during \gc are eliminated,} because they would require significant
device I/O. To achieve this, \name{} organizes H2 into regions with
similar-lifetime objects and deals differently with liveness analysis
and space reclamation. 
For liveness analysis, \name{} identifies live H2 regions by
tracking forward (H1 to H2) and cross-region (in H2) references during
\gc. To identify live objects in H1, \name{} explicitly tracks
backward references (H2 to H1) and fences \gc scans in H2. \name{}
tracks backward references using a card table optimized for
storage-backed heaps, minimizing I/O traffic to the underlying device
during \gc. 
For reclamation, the collector reclaims H1 objects as usual. For H2
regions, unlike existing region-based
allocators~\cite{gog:broom,khanh:yak} \name{} resolves the
space-performance tradeoff for reclaiming space differently. Existing
allocators reclaim region space eagerly by moving live objects to
another region, which would generate excessive I/O for storage-backed
regions. Instead, \name{} uses the high capacity of NVMe SSDs to
reclaim entire regions lazily, avoiding slow object compaction on the
storage device.

\paragraph{Applying \name{}:}
Big data frameworks exhibit significant  diversity with respect to the
objects they move off-heap. We investigate how Spark and Giraph, two
widely used frameworks, resolve the trade-off between GC cost due to
large heaps and the overhead of off-heap accesses. Spark users
explicitly store immutable cached data on the device, while Giraph
transparently (without user hints) offloads mutable objects to the
device.  We modify both frameworks to use \name{} for two very
different purposes.

We implement \name{} and its mechanisms in OpenJDK, extending the
Parallel Scavenge garbage collector. We also extend the interpreter
and the C1 and C2 Just-in-Time (JIT) compilers to support object
updates in H2 during application execution.
Our evaluation shows that \name{} improves performance by up to 73\%
and 28\% compared to the native Spark and Giraph, respectively.
\name{} provides similar or better performance by consuming up to
$8\times$ and $1.2\times$ less DRAM capacity than native Spark and
Giraph, respectively. Also, it outperforms
Panthera~\cite{wang:Panthera}, a garbage collector specialized for
hybrid memories, by up to 69\%.

Overall, our work makes the following contributions:

\begin{itemize}
	\item
        We introduce a dual heap approach to reduce \ser and memory
        pressure in big data frameworks, by adding a second,
        high-capacity, managed heap over a fast storage device.
	\item
	      We propose a hint-based interface based on \emph{key-object
		      opportunism} that enables frameworks to mark candidate objects
	      in a coarse-grain manner and select when to move them to the
	      second heap.
	\item
        We show the applicability of \name{} as: (1) a large, on-heap,
        compute cache in Spark to store intermediate results, and (2)
        a high-capacity heap in Giraph to store messages and edges.

\end{itemize}

\section{Background}

This section provides background related to  JVM garbage collection
and serialization/deserialization. 

\paragraph{Garbage Collection:}
Modern collectors exploit the generational hypothesis that many
objects die young. For this reason, they divide the managed heap into
a young generation for new objects and an old generation for objects
that survive multiple young (minor) collections~\cite{Ungar:1984:GSN}.
They further divide the young generation into an eden space and two
survivor spaces, called \emph{from-space} and \emph{to-space}.
Application (mutator) threads allocate new objects into the eden
space. When the eden space becomes full, garbage collectors perform a
minor \gc. During minor \gc, the garbage collector identifies live
objects in the eden space and \emph{from-space}. Then, it moves live
objects to the \emph{to-space} and the mature objects to the old
generation. When the managed heap becomes full, the JVM performs a
full (major) \gc, which scans and compacts both old and young
generations.

Although JVMs used to support only DRAM-resident managed heaps, today,
they can allocate either the entire heap or the old generation over a
storage device using memory-mapped I/O (e.g., Linux \emph{mmap}).
However, existing garbage collectors are tuned for DRAM-backed heaps,
increasing the collection overhead drastically for storage-backed
heaps~\cite{nvm-gc-eurosys}. Their design targets DRAM, which provides
low latency and high throughput regardless of operation types
(read/write) and access patterns (random/sequential). On the other
hand, block-addressable storage devices (e.g., NVMe SSDs) exhibit
higher latency and lower throughput than DRAM. 

\paragraph{Object Serialization:}
Java serialization enables the conversion of a memory-resident object
into a form that is convenient for transportation off-heap (memory,
storage, or network) and can even be shared across JVMs.
\emph{Serialization} transforms Java objects in the managed heap into
a byte stream, and \emph{deserialization} reconstructs the Java
objects from byte streams into heap representations (with references).
During \ser, the serializer traverses the object graph to identify all
objects that need to be serialized, starting from the root object
selected for off-heap placement. When serializing an object, the
serializer omits fields marked with the \emph{transient} modifier.
Transient fields are initialized to a default value during
deserialization based on the serializer implementation.

Java serialization is a complex process that introduces significant
limitations and overheads during execution. Serialization limits the
objects that can be moved off-heap, as it requires self-contained
entities without references to and from the managed heap, i.e., only
serializable objects~\cite{grochowski2019serialization,
niemeyer2005learning, harold2006java}. In addition, extracting and
recreating the object state requires mechanisms that bypass
constructors and ignore class and field accessibility.
Performance-wise, traversing the object graph requires effort
proportional to the volume of objects in the transitive closure of the
root object. Most relevant to our work, \ser generates many temporary
objects while transforming objects into byte streams and vice-versa.
Temporary objects put more pressure on the heap and lead to more
frequent \gc cycles. Recent work identifies \ser as a significant
performance bottleneck in big data analytics
frameworks~\cite{navasca:gerenuk, nguyen:skyway, matei:champions,
taranov:naos}.

\section{\name{} Design}
\subsection{Overview}
\label{sec:over}

The key idea of \name{} is to extend the JVM to use a second,
high-capacity managed heap (H2) over a fast storage device that
coexists with the regular managed heap (H1). \name{} manages the two
heaps differently and hides their heterogeneity, providing the
abstraction of a large homogeneous managed heap to big data
applications. We design \name{} based on our observations about
objects and their management in big data analytics frameworks.

\paragraph{Which objects to move to H2 and when?} 
We observe that different managed big data frameworks maintain
off-heap stores to move specific objects that are long-lived and are
reused across computation stages. Also, these objects differ in their
update patterns. For example, Spark only moves immutable objects
off-heap, Giraph moves objects that are \emph{eventually} immutable.
Such objects are those that, once immutable, are reused by later
computation stages. 

We provide a novel hint-based interface based on key-object
opportunism~\cite{hayes:key_opport} to identify specific objects to
move to H2. These objects are ones that frameworks move off-heap.
Frameworks use our interface to (1) tag with a label the root key
object appropriate for placement in H2 and (2) advise \name{} when to
move objects in H2. Decoupling the selection of the candidate objects
from their transfer to H2 enables \name{} to be framework agnostic.
These hints are translated into two native function calls at runtime.
Our hint-based interface works at the framework level and is
transparent to applications written on top of such frameworks,
requiring minimal user effort. We provide more details in
\S\ref{sec:api}.

\paragraph{How to reclaim dead objects in H2 without \gc scans?}
Scanning the storage-backed H2 for liveness analysis and compacting
objects for space reclamation incurs a high \gc overhead due to
excessive device I/O traffic. \name{} reduces the high \gc overhead by
organizing H2 in virtual memory as a region-based heap. Each region
hosts object groups with similar lifetimes to reclaim dead objects in
bulk. We observe that analytics frameworks, such as Spark and Giraph,
organize groups of objects in data structures, such as array with a
single-entry root reference (key objects). Most objects that are
reachable by root key-objects exhibit a similar lifetime. 

\name{} leverages the high capacity of NVMe SSDs to resolve the
space-performance tradeoff differently than existing
work~\cite{khanh:yak, gog:broom}. Current works target DRAM-backed
heaps and focuses on freeing address space eagerly by scanning regions
and moving live objects with cross-region references to other regions.
However, \name{} reclaims H2 space lazily with low overhead by freeing
whole regions and their objects in bulk. To ensure memory safety while
reclaiming dead H2 regions, \name{} must take into account forward (H1
to H2) and cross-region references (details in \S\ref{sec:free}).

\begin{figure}[t]
	\centering
  \includegraphics[width=\linewidth]{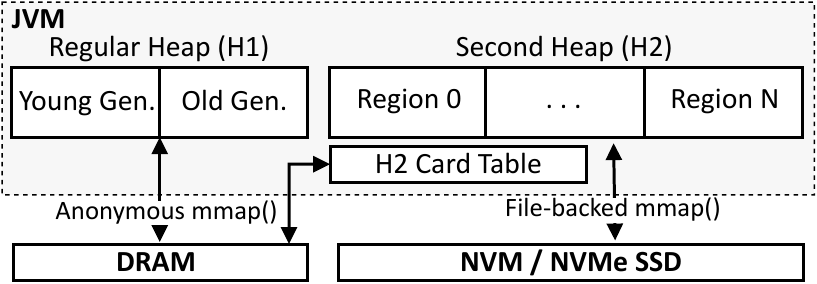}
  \caption{\name{} design overview.}
	\label{fig:teraheap}
\end{figure}

\paragraph{\name{} architecture:} 
At a high level, \name{} uses two heaps as shown in
Figure~\ref{fig:teraheap}: regular heap (H1) and second, high-capacity
heap (H2). Unlike DRAM-backed H1, H2 is memory-mapped over a storage
device, allowing direct access to deserialized objects without \ser.
Memory-mapped I/O eliminates the need to use a custom reference lookup
mechanism in the JVM to identify objects on the device, as the OS
virtual memory mechanism performs this translation. Also, to identify
references from H2 to H1, \name{} uses a card table optimized
for storage-backed heaps (details in \S\ref{sec:back_ref}).

We have designed \name{} to be agnostic to the specific device that
backs H2. However, the intention is to map H2 over fast storage
devices, either block-addressable NVMe SSDs or byte-addressable NVM.
Such devices are amenable to memory mapped I/O due to their high
throughput and low latency for small request sizes (4\:KB) regardless
of the access pattern~\cite{apapag:fmap}. NVMe SSDs are particularly
attractive as datasets grow because they provide high density
(capacity) and lower cost per bit compared to DRAM and
NVM~\cite{weiner:tmo}. 

Next we discuss how \name{} solves the three main challenges related
to:
(1) identifying and moving candidate objects to H2,
(2) reclaiming dead objects in H2 without \gc scans and I/O traffic,
and
(3) tracking backward references (H2 to H1) with low \gc cost and I/O
overhead.

\subsection{Identifying and Moving Candidate Objects to H2}
\label{sec:api}

\name{} provides a hint-based interface, enabling frameworks to tag
root key-objects with a label for H2 movement. For the tagging
operation of H2 candidate objects, we add a new field (eight bytes for
alignment purposes) in the Java object header. Avoiding the extra
field requires additional JVM book-keeping and meta-data, increasing
\gc time. The \name{} interface consists of the following function
calls.

\paragraph{h2\_tag\_root(obj, label):} 
The framework uses \texttt{h2\_tag\_root()} to tag a root key-object
with a label.

\paragraph{h2\_move(label):} 
The framework uses \texttt{h2\_move()} to advise \name{} to move all
objects with specified label to H2.
During the next major
\gc, the garbage collector marks objects in the transitive closure of
the root key-object with the label. Typically, frameworks
can use \texttt{h2\_move()} once their object group becomes immutable,
however, immutability is not a strict requirement for movement to H2
and partly depends on storage device
characteristics~\cite{swanson:nvm}. For instance, in Spark, all
objects can be moved when marking the root key-object, whereas, in
Giraph, objects are best moved at the end of each computation stage,
possibly much later than when marking the root key-object.

Delaying the move to H2 runs the danger of creating out-of-memory
errors because H1 may fill before \texttt{h2\_move()} is called.  To
avoid this, \name{} monitors the space that live objects occupy at the
end of each major \gc. If the live objects occupy more space in H1
than a high threshold (e.g., 85\% of H1), \name{} will move marked
objects to H2 during the next major \gc without waiting for
\texttt{h2\_move().} 

At this point, if \name{} moves all marked objects to H2, it may incur
excessive device traffic, e.g., in case some of these objects may be
updated frequently prior to the application using \texttt{h2\_move().}
To mitigate this effect, \name{} uses a low threshold mechanism as
well, which limits how many marked objects will move to H2 when
\name{} detects high H1 pressure prior to seeing an
\texttt{h2\_move()} hint. In our evaluation, we examine the
alternative of not using the \texttt{h2\_move()} and relying only on
the high-low threshold mechanism.

\name{} moves all objects with the same label in the same H2 region
until it exhausts the region space to reclaim them en masse. However,
the transitive closure might include JVM metadata objects and
specialized objects which have a longer lifetime than the rest of the
objects in the closure. These objects can extend the region's
lifetime, preventing free operations in H2. Thus, \name{} excludes
from moving to H2: 
(1) JVM metadata from the transitive closure, such as \emph{class
objects}~\cite{ClassJav88:online} and the class loader, and 
(2) specialized objects that inherit \emph{java.lang.ref.Reference}
class~\cite{Referenc77:online}.

\name{} moves marked objects from H1 to H2 during major \gc.  The main
overhead of \name{} for major \gc is the transfer of objects from H1
to H2. To reduce this cost, \name{} uses explicit asynchronous I/O. We
avoid multiple system calls for small-sized objects ( <1\:MB), using a
promotion 2\:MB buffer per region in H2 that writes objects to the
device in batches.

\subsection{Reclaiming Dead Regions}
\label{sec:free}
\begin{figure}
	\includegraphics[width=\linewidth]{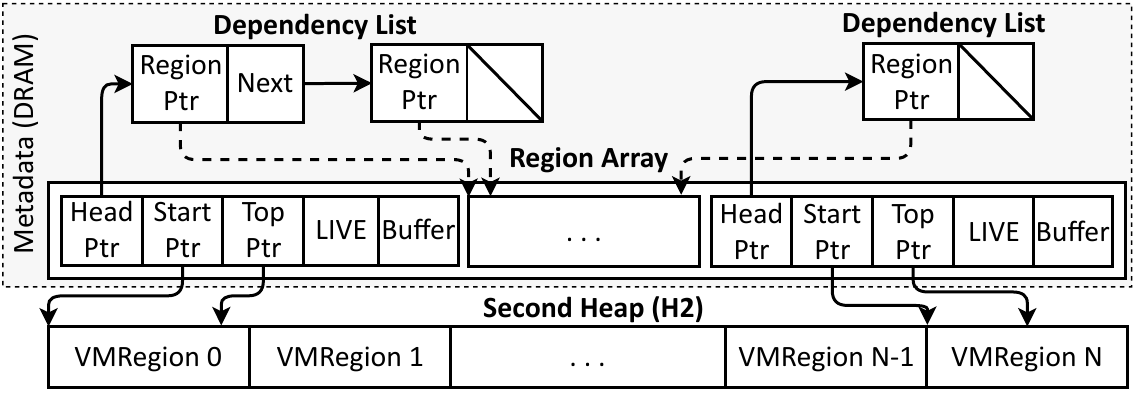}
	\caption{H2 allocator metadata in DRAM (Ptr=Pointer).}
	\label{fig:allocator}
\end{figure}

Figure~\ref{fig:allocator} shows the region-based organization of H2
in virtual memory and each region metadata in DRAM. We do not impose
any restrictions on regions, allowing objects in any region to refer
to each other. \name{} ensures that while reclaiming a region, none of
the objects in the region are referenced from live H1 objects or live
H2 objects in other regions. To find such regions, \name{} tracks
cross-region and forward references without scanning H2 objects, which
would generate excessive I/O.
 
\paragraph{Cross-region references in H2:}
To allow internal H2 references \emph{across} regions, \name{} tracks
the direction of cross-region references. As shown in
Figure~\ref{fig:allocator}, \name{} keeps a dependency list in
per-region metadata in DRAM. Each node of the dependency list points
to a (different) region referenced by objects of the current region.
When we move objects to H2 we check if they have references in
existing H2 regions. Then, the H2 allocator adds a new node (if it
does not exist) to the dependency list of the region where objects
will be moved. The size of dependency lists is small, on average 10
nodes per region in our evaluation.

We also explore a simpler, \emph{Union-Find} approach, using the
notion of region groups that avoids tracking the direction of
cross-region references. We track cross-region references by logically
merging the source and destination regions in a single region group.
Region groups grow over time to include all regions with cross-region
references. If there is any reference from H1 to any object in the
group's regions, then we consider the group alive. This approach does
not consider the direction of region references, missing the
opportunities to reclaim dead regions with no incoming references. For
instance, if there is a reference from region X to Y and a reference
from region Y to Z, all three regions belong to the same region group
and can be reclaimed when the whole group dies. We find that the
direction of references matters for more efficient space reclaiming.
In the previous example, if only region Z is referenced by H1, then
regions X and Y can still be reclaimed.

\paragraph{Forward references (H1 to H2):}
\name{} avoids scanning H2 objects by fencing the garbage collector
from crossing into H2 from H1. This requires identifying all
references from H1 to H2 and marking the referenced H2 objects as
alive. \name{} uses a LIVE bit in the per-region metadata
(Figure~\ref{fig:allocator}) that signifies the objects in the region
are reachable from H1. The garbage collector clears LIVE bits at the
beginning of the major \gc. Upon encountering a reference from an
object in H1 to an object in H2, the collector sets the corresponding
region bit. If the dependency list of the current region is not empty,
then we traverse the dependency lists of each dependent region
recursively, setting their LIVE bits, as well.

\paragraph{Freeing dead regions:}
At the end of major \gc{}, any H2 region not marked as LIVE is not
reachable from any H1 object nor any H2 regions. To free these dead
regions, we set their allocation pointer to zero, and delete their
dependency list (Figure~\ref{fig:allocator}). We note that upon JVM
shutdown, we free all H2 metadata in DRAM.

\subsection{Tracking Backward References (H2 to H1)}
\label{sec:back_ref}
Fencing \gc scans in H2 further requires tracking backward
references from H2 to H1, as the garbage collector must not reclaim H1
objects referenced by live H2 objects. The key difficulty is that H2
objects can reference objects in both H1 generations and need to be
tracked differently. Young objects in H1 change location during minor
\gc while old objects move only during major \gc. 

Scanning H2 to identify backward references may incur significant
overhead, depending on the size of H2 and its backing device. Instead,
we use an extended card table for H2, optimized for use with storage
devices. The H2 card table is a byte array (in DRAM) with one byte per
fixed-size H2 segment (similar to vanilla JVM). Although using a
\emph{remembered set} provides more precise information about backward
references, it increases memory consumption for regions with many
references, especially as H2 size grow with storage device capacity.
It also requires a more elaborate and expensive post-write
barrier~\cite{detlefs:remset}.

\paragraph{Setting H2 card states:}
We expect H2 to be much larger than H1. Thus, we increase the size of
H2 card segments to reduce the number of cards and the card scanning
overhead during collections. However, larger card segments require
scanning more objects, in case they are dirty, introducing device I/O.
To reduce the number of objects scanned during minor \gc, we avoid
scanning H2 objects that only reference objects in the old generation
of H1. In minor \gc, the collector does not move or reclaim objects in
the old generation. Thus, we design an H2 card table where each card
entry is in one of four states:
(1) \emph{clean}, when there are no backward references,
(2) \emph{dirty}, indicating object update by mutator threads,
(3) \emph{youngGen}, indicating references only to the young
generation, or
(4) \emph{oldGen}, indicating references only to the old generation.

When an application thread updates an H2 object, \name{} marks the
corresponding H2 card as \emph{dirty} in the post-write barrier.
During \gc, we change the card value from \emph{dirty} to
\emph{OldGen} if objects in the dirty card segment only reference
objects in the old generation. Otherwise, we change the card value to
\emph{YoungGen}. We set the card value as \emph{clean} only if there
are no backward references in the card segment. In minor \gc, we only
scan the objects in the card segments whose cards are marked as
\emph{dirty} or \emph{youngGen}. In major \gc we also scan
\emph{oldGen} objects. We adjust all backward references in both minor
and major \gc to refer to the new H1 object locations.

\paragraph{Scanning H2 card table:}
\gc is multithreaded, and therefore, the H2 card table must support
concurrent access from multiple threads without synchronization. We
divide H2 into slices to avoid contention between \gc threads. Each
slice contains a number of fixed-size stripes equal to the number of
\gc threads. Each \gc thread processes the stripe with the same Id
within each slice. Therefore, each \gc thread operates on the same
stripe Id in all H2 slices.

We have to solve the access to the boundary (first and last) cards in
each stripe. Objects may span card segments and stripe boundaries.
Given that a separate \gc thread processes each stripe, two threads
may need to access each boundary card.
To avoid thread synchronization, the garbage collector can avoid
cleaning the boundary cards. If boundary cards become dirty, they can
remain dirty throughout execution. However, in every \gc, the garbage
collector should scan the corresponding card segments for objects with
backward references. This drawback is a significant issue for H2
because (1) we use large card segments to reduce H2 card table size,
and (2) the card segments are mapped to a storage device, which
results in high I/O traffic when scanning objects.

\name{} resolves the problem of the dirty boundary cards by aligning
objects to stripes and guaranteeing that no two threads will need to
access the same card. \name{} uses a larger stripe size equal to the
H2 region size because the \name{} guarantees that objects do not span
H2 regions.

\section{\name{} for Parallel Scavenge GC}

We implement \name{} in OpenJDK8 (jdk8u345) which is a
long-term-support (LTS) version, extending the Parallel Scavenge (PS)
garbage collector and export \name{}'s interface through the
\emph{sun.misc.Unsafe} class to frameworks. PS is a generational
garbage collector which divides its heap into young and old
generations. Next, we discuss our extensions in (1) the post-write
barriers in interpreter and just-in-time (JIT) compilers, (2) minor
\gc, and (3) major \gc.

\paragraph{Post-write barriers:}
PS uses a post-write barrier and a card table to track updates in old
generation objects that generate references to young objects. Such
updates may originate from interpreted or JIT compiled methods with
the C1 and C2 JVM compilers. When a mutator thread updates an object
in the old generation, this operation is followed by the post-write
barrier that updates the corresponding entry in the H1 card table. 

To examine if the mutator thread updates an object that belongs to H1
or H2, we use an additional range check in the post-write barrier.
This range check selects the appropriate (H1 or H2) card table, which
we then mark with the existing post-write barrier code. 
We extend post-write barriers by augmenting the template-based
interpreter and the JIT compilers to generate assembly code for the
necessary checks, guarded by the \emph{EnableTeraHeap} flag. We
evaluate the overhead of our modifications to post-write barriers
using the DaCapo benchmark suite~\cite{blackburn:dacapo}. The overhead
is small and within 3\% over total execution time on average across
all benchmarks. The extra overhead is zero for applications that do
not set \emph{EnableTeraHeap}.

\paragraph{Minor \gc:}
In minor \gc, we perform two key tasks during liveness analysis: (1)
fence PS from scanning objects in H2 and (2) prevent reclamation of
H1 objects referred from H2 objects (backward references). For the
first task, we introduce an additional reference check in the liveness
analysis to fence PS from scanning references that cross from H1 to
H2. We scan the H2 card table for the second task to identify and
update backward references and the H2 card state.

\paragraph{Major \gc:}
The major \gc in PS is divided into four main phases. In the first
phase (\emph{marking}), PS recursively scans both generations starting
from roots (e.g., thread stacks, global variables, registers) and
marks live objects. We extend the \emph{marking} phase to perform five
extra tasks. At the beginning of the marking phase, we reset all LIVE
bits of the H2 regions metadata. We mark all objects in H1 that are
referenced by H2 as live. We add a reference range check (similar to
minor \gc) that detects forward references (H1 to H2) to fence the PS
from scanning objects in H2 and sets the LIVE bit of the corresponding
region. We identify the root objects tagged with a label through
\name{} interface and calculate their transitive closure. At the end,
we free all dead regions in H2.

PS assigns a new memory location to each live object in the second
phase of major \gc (\emph{pre-compaction}). We extend this phase to
identify which objects discovered in the \emph{marking} phase should
be moved in H2. We assign these objects an address from H2 using their
label. 

During the third phase (\emph{pointer-adjustment}), PS adjusts the
references of each object to point to new object locations. We extend
this phase to perform two tasks, update cross-region and backward
references. We detect cross-region references when we adjust the
references of newly marked H1 objects that are candidates for H2
transfer by checking if they reference existing H2 regions. This task
is performed solely when the objects are still in DRAM.  \name{} scans
objects for backward references also when they reside in DRAM and
before they are moved to H2. If we detect an object with backward
reference, we mark only the corresponding H2 card as dirty. 
Finally, in the fourth phase (\emph{compaction}), PS moves objects to
their new locations in H2.

\section{Applications of \name{}}

In this section, we describe how we use \name{} in two widely used
frameworks, Spark~\cite{Zaharia:Spark} and Giraph~\cite{giraph}. Note
that Spark and Giraph differ significantly in how they use off-heap
memory. Spark uses off-heap memory to cache intermediate results,
avoiding expensive recomputation. Cached objects are immutable at
allocation time. On the other hand, Giraph offloads mutable objects,
i.e., vertices, edges, and messages, to off-heap memory to ensure
adequate DRAM is available for each superstep. Giraph updates vertex
values throughout the computation, whereas edges and messages become
immutable after graph loading (edges) or at the end of a superstep
(messages).

Spark users explicitly annotate objects that need to be moved off-heap
with the \texttt{persist()} call. Giraph transparently selects and
moves objects to the storage device without application interaction.
It maintains an \emph{out-of-core} scheduler that monitors memory
pressure in the managed heap and decides which vertices, edges, and
messages to move off-heap. The out-of-core scheduler selects based on
a Least Recently Used (LRU) policy which objects to move off-heap.

Spark maintains deserialized objects in memory; that incurs significant
\ser overhead during the off-heap movement. Giraph tries to reduce memory
consumption on the managed heap and serializes
vertices, edges, and messages into byte arrays, at allocation time.
Therefore, Giraph does not require \ser when moving these byte arrays
off-heap on the storage device.
Next, we discuss how we extend Spark and Giraph to use \name{}.

\begin{figure}[t]
	\centering
	\includegraphics[width=0.95\linewidth]{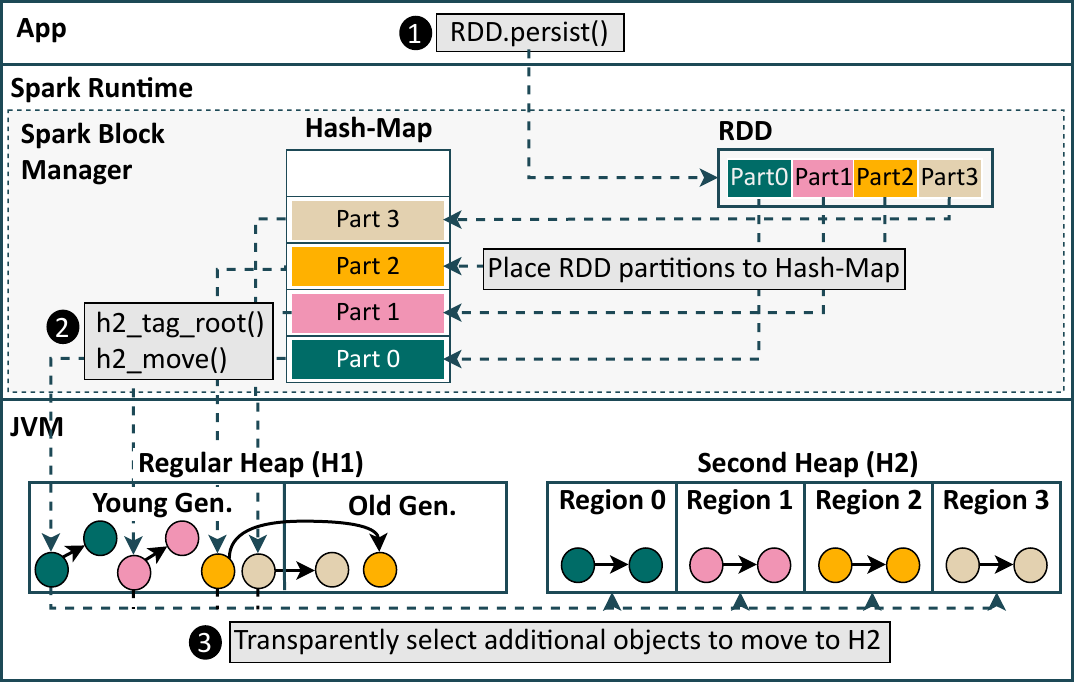}
	\caption{Spark interacting with \name{} interface.}
	\label{fig:spark_diag}
\end{figure}

\paragraph{Spark:}
Each spark executor logically divides its memory into two main spaces,
execution memory for computation and storage memory for on-heap
caching of intermediate results.  Spark abstracts intermediate results
as immutable collections using three sets of APIs~\cite{spark:apis}:
resilient distributed datasets (RDDs)~\cite{rdd}, dataframes, and
datasets.

Spark requires only slight modifications to use \name{}. In Spark we
mark all cached partitions of RDDs, Dataframes, or Datasets, as root
objects for moving to H2. Figure~\ref{fig:spark_diag} shows the flow
of Spark caching operations using \name{}:
\circled{1} The application code invokes \texttt{persist()} without any
modifications.
\circled{2} The Spark block manager places the selected data in the
compute cache, a hashmap that contains all cached partitions. The
Spark block manager caches each partition independently, maintaining
per-partition entries in the hashmap. When the Spark block manager
stores a new partition in the hashmap, we mark the partition
descriptor as a root key-object with the \texttt{h2\_tag\_root()},
providing as label the RDD, dataset, or dataframe Id. At the same
time, we advise JVM to move marked partitions to H2, using
\texttt{h2\_move()}.
\circled{3} \name{} transparently marks additional objects and moves
them to H2 during the major \gc.

\begin{figure}[t]
	\centering
	\includegraphics[width=0.95\linewidth]{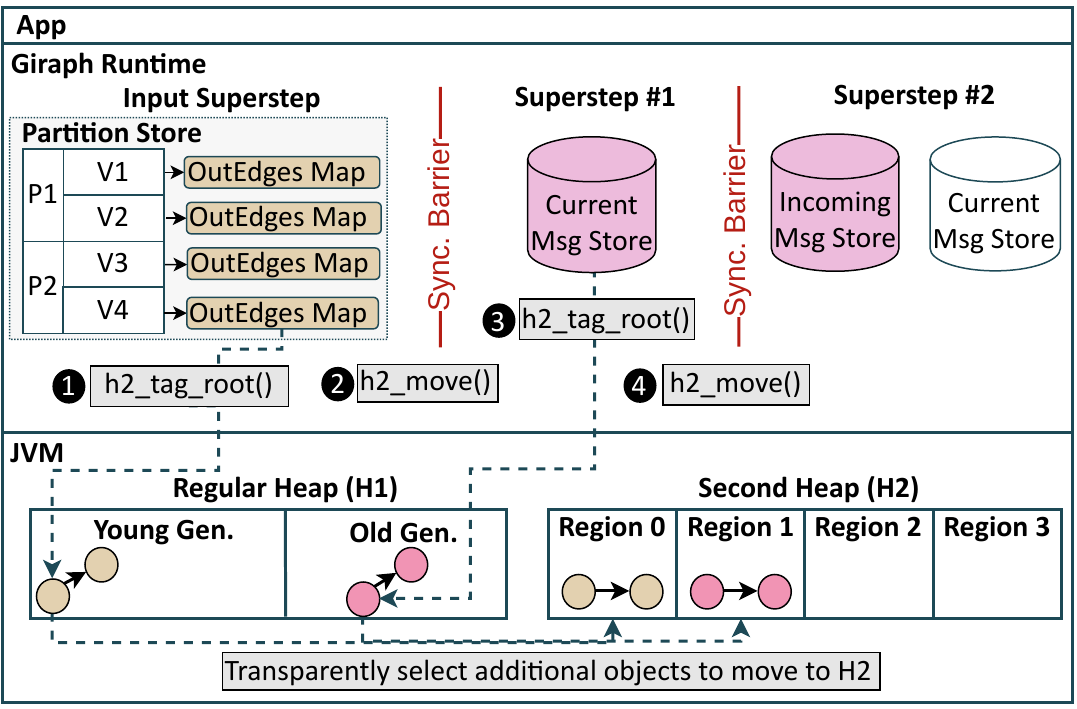}
	\caption{Giraph interacting with \name{} interface.}
	\label{fig:giraph}
\end{figure}

\paragraph{Giraph:}
Giraph computes in supersteps, with a synchronization barrier between
supersteps. It loads and partitions the graph during the input
superstep.  A graph partition organizes its vertices in a hashmap,
with each vertex belonging to a single partition. Each vertex
maintains a map containing its outgoing edges. In each superstep, each
vertex consumes all of its incoming messages from the previous
superstep and updates its value.  Then, it sends its updated value to
its outgoing edges in a new message (per vertex). Messages produced in
the current superstep are only consumed in the next superstep.
Messages become immutable at the end of each superstep after the
coordination phase guarantees they have been received and saved
completely.  Thus, in each superstep, Giraph has two message stores:
the \emph{incoming message store} with messages from the previous
superstep (immutable) and the \emph{current message store} with
messages of the current superstep (mutable).

To use \name, Giraph requires small modifications, as well. We extend
Giraph by marking edges and incoming messages as root objects. We do
not mark vertices because they have frequent updates, and they will
increase device (write) traffic. Note that edges and messages
constitute a large portion of the heap~\cite{giraph}.
Figure~\ref{fig:giraph} shows the  flow of Giraph execution using
\name{}:
\circled{1} When Giraph loads a new vertex at the input superstep, it
marks the vertex's map that contains the outgoing edges with
\texttt{h2\_tag\_root()}, providing the superstep id as label.
\circled{2} At the end of the input superstep, Giraph advises \name{}
to move marked edges to H2, using \texttt{h2\_move()} in the next major
\gc.
\circled{3} In each superstep, Giraph marks the generated messages of
the  \emph{current message store} with \texttt{h2\_tag\_root()},
providing as label the superstep id.
\circled{4} At the beginning of each (next) superstep, Giraph advises
\name{} to move in H2 all marked messages from the previous superstep
in the next major \gc, using \texttt{h2\_move()}.

\section{Experimental Methodology}
\label{sec:method}

\setlength\heavyrulewidth{1pt}
\begin{table}[t]
  \centering
  \resizebox{\columnwidth}{!}{%
    \begin{tabular}{llclc}
      \multicolumn{1}{c}{\textbf{ID}} & \multicolumn{1}{c}{\textbf{CPU}}  & \multicolumn{1}{c}{\textbf{\begin{tabular}[c]{@{}c@{}}DRAM\end{tabular}}} & \multicolumn{1}{c}{\textbf{Device}}                                                & \multicolumn{1}{c}{\textbf{Kernel}} \\
        \toprule
        \begin{tabular}[c]{@{}l@{}}NVMe \\ Server\end{tabular}  & \begin{tabular}[c]{@{}l@{}}Intel Xeon E5-2630 \\ 32 Cores @  2.4 GHz\end{tabular} & 256\:GB & \begin{tabular}[c]{@{}l@{}}2 TB Samsung PM983 \\ PCI Express NVMe SSD\end{tabular} & 4.14 \\ \hline
        \begin{tabular}[c]{@{}l@{}}NVM \\ Server\end{tabular} & \begin{tabular}[c]{@{}l@{}}Intel Xeon Platinum \\ 24 cores @ 2.4 GHz\end{tabular} & 192\:GB & \begin{tabular}[c]{@{}l@{}}3 TB Intel Optane DC \\ Persistent Memory\end{tabular}  & 3.10 \\ \bottomrule

    \end{tabular}%
    }
    \caption{NVMe and NVM servers properties.}
    \label{tab:server}
\end{table}

We answer the following questions in our evaluation: 
\begin{enumerate}
  \item How does \name{} perform compared to native JVM and
    state-of-the-art Panthera with NVMe SSD and NVM?
  \item What are the space requirements of H2? 
  \item What is the overhead of tracking references and moving
    objects in H2 during \gc?
  \item How does \name{} scale with increasing numbers of mutator
    threads and dataset sizes?
\end{enumerate}

\paragraph{Server infrastructure:}
We evaluate \name{} both with block-addressable NVMe SSDs and
byte-addressable NVM as the backing device for H2.
Table~\ref{tab:server} shows the properties of each server. Our NVM
server operates in two modes: 
(1) \emph{App Direct Mode} uses 192\:GB DRAM as main memory and 2\:TB
NVM as persistent storage device.
(2) \emph{Mixed Mode} partitions NVM to use 1\:TB in memory mode and
2\:TB in App Direct mode. DRAM (192\:GB) acts as a cache for 1\:TB NVM
controlled by the CPU's memory controller. In AppDirect mode, the
system mounts NVM on an ext4-DAX file system to establish direct
mappings to the device.

\paragraph{Baseline and \name{} configurations:}
We run Spark with OpenJDK8, OpenJDK11, and OpenJDK17. We run Giraph
v1.2 with Hadoop v2.4 and OpenJDK8, as it does not support more recent
versions of OpenJDK. We use two garbage collectors in different
configurations: PS in OpenJDK8 and OpenJDK11, and Garbage First (G1)
in OpenJDK17.
We use an executor with eight mutator threads for both Spark and
Giraph. For PS, we use 16 \gc threads for minor \gc and the default
single-threaded old generation \gc. 
G1 uses two parameters: (1) the number of parallel \gc threads, which
we set to eight (max value), and (2) the number of concurrent (to
mutator) threads, which we set to two as the recommended
configuration~\cite{g1:manual} is one-fourth of the parallel \gc
threads.

\begin{table}[t]
\centering
\resizebox{\columnwidth}{!}{%
\begin{tabular}{l|cccc}
\multicolumn{1}{c|}{\textbf{Baseline}} &
  \textbf{DRAM} &
  \textbf{\begin{tabular}[c]{@{}c@{}}NVMe SSD\end{tabular}} &
  \textbf{\begin{tabular}[c]{@{}c@{}}NVM\\ (App Direct Mode)\end{tabular}} &
  \textbf{\begin{tabular}[c]{@{}c@{}}NVM \\ (Memory Mode)\end{tabular}} \\
    \toprule
    Spark-SD   & Heap & Off-heap & -        & -    \\
    Spark-SD   & Heap & -        & Off-heap & -    \\
    Spark-MO   & -    & -        & -        & Heap \\
    Giraph-OOC & Heap & Off-Heap & -        & -    \\
\bottomrule
\end{tabular}%
}
\caption{Summary of baselines.}
\label{tab:baselines}
\end{table}

Table~\ref{tab:baselines} summarizes the Spark and Giraph
configurations we use as baselines.
The two Spark-SD configurations place executor memory (heap) in DRAM
and cache RDDs in the on-heap cache, up to 50\% of the total heap
size. Any remaining RDDs are serialized in the off-heap cache, over
either NVMe SSD (first line of Table~\ref{tab:baselines}) or NVM in
App Direct Mode (second line of Table~\ref{tab:baselines}).
Spark-MO places executor memory (heap) over NVM in Memory Mode, caching
all RDDs on-heap.
Giraph-OOC places the heap in DRAM and offloads vertices, edges, and
messages off-heap to the NVMe SSD.

We configure \name{} to allocate H1 on DRAM and H2 over a file in NVMe
SSD or NVM via memory-mapped I/O (\mmio). The file in both NVMe and
NVM servers is mapped to the JVM virtual address space where the
application can access the data with regular load/store
instructions~\cite{apapag:fmap}. Our experiments show that machine
learning (ML) workloads in Spark access the individual elements of
cached RDD partitions sequentially. For this reason, we configure
\name{} for Spark ML workloads to use huge pages (2\:MB) in H2 to
reduce the frequency of page faults. Instead of the native
\emph{mmap}, we use HugeMap~\cite{jmal:hugemap} a custom, open source,
\mmio path that enables huge pages for file-backed mappings.

\renewcommand{\arraystretch}{1.1}
\begin{table}[t]
	\centering
	\resizebox{\columnwidth}{!}{%
		\begin{tabular}{p{0.01cm}l|ccccc}
			                                                                 &
			\multicolumn{1}{c|}{}
      & \textbf{\begin{tabular}[c]{@{}c@{}}NVMe \\
        Server\end{tabular}}        &
        \textbf{\begin{tabular}[c]{@{}c@{}}Data-\\set\end{tabular}} &
          \textbf{\begin{tabular}[c]{@{}c@{}}Spark-\\SD\end{tabular}}
            &
            \textbf{\begin{tabular}[c]{@{}c@{}}Spark-\\MO\end{tabular}} & \textbf{\begin{tabular}[c]{@{}c@{}}Tera\\Heap\end{tabular}}      \\
			                                                                 & \multicolumn{1}{r|}{\textbf{GB}}                                       &
			\multicolumn{1}{l}{\textbf{DRAM}}                                & \multicolumn{1}{c}{\textbf{Size}}                                      & \multicolumn{1}{c}{\textbf{Heap}}                             & \multicolumn{1}{c}{\textbf{Heap}}                          & \multicolumn{1}{c}{\textbf{H1}}                                                  \\
			\toprule
      \multirow{5}{*}{\rotatebox[origin=c]{90}{\textbf{GraphX}}}       
      & PageRank (PR)
      & 80                                                          
      & 32                                                             
      & 64                                                          
      & 1024              
      & 64
      \\

      &\begin{tabular}[c]{@{}l@{}} Connected Components (CC)\end{tabular}
      & 84
      & 32
      & 68
      & 1024
      & 68
      \\

      & Shortest Path (SSSP)
      & 58
      & 32
      & 42
      & 650
      & 42 \\

      &
                                                                       SVDPlusPlus (SVD)
                                                                       &
                                                                       40
                                                                       &
                                                                       2
                                                                       &
                                                                       24
                                                                       &
                                                                       500
                                                                       &
                                                                       24
                                                                       \\

                                                                       &
                                                                       Triangle
                                                                       Counts
                                                                       (TR)
                                                                       &
                                                                       80
                                                                       &
                                                                       2
                                                                       &
                                                                       64
                                                                       &
                                                                       64
                                                                       &
                                                                       64
                                                                       \\
                                                                       \bottomrule
                                                                       \multirow{4}{*}[-2.5pt]{\rotatebox[origin=c]{90}{\textbf{MLlib}}}
                                                                       &
                                                                       Linear
                                                                       Regression
                                                                       (LR)
                                                                       &
                                                                       70
                                                                       &
                                                                       256
                                                                       &
                                                                       54
                                                                       &
                                                                       1084
                                                                       &
                                                                       54
                                                                       \\

			                                                                 & Logistic Regression (LgR)                                                    & 70                                                          & 256                                                            & 54                                                          & 1084              & 54 \\

                                                                       & \begin{tabular}[c]{@{}l@{}}Support Vector Machine (SVM)\end{tabular}
                                                                       & 48                                                          
                                                                       & 256                                                            
                                                                       & 32                                                          
                                                                       & 620               
                                                                       & 32 \\

			                                                                 & \begin{tabular}[c]{@{}l@{}}Naive Bayes Classifier (BC)\end{tabular}
                                                                         & 98 
                                                                         & 21
                                                                         & 82
                                                                         & 82
                                                                         & 82 \\
			\bottomrule
			\rotatebox[origin=c]{90}{\textbf{SQL}}
      & RDD-RL
      & 63
      & 16
      & 47
      & 96
      & 47 \\
			\bottomrule
		\end{tabular}%
	}
  \caption{Configuration for each workload on NVMe and NVM servers for
  Spark-SD, Spark-MO, and \name{}.} 
  \label{tab:spark_conf}
\end{table}

\begin{table}[t]
	\centering
	\resizebox{\columnwidth}{!}{%
		\begin{tabular}{l|cccccc}
			\multicolumn{1}{c|}{}            &
      \textbf{\begin{tabular}[c]{@{}c@{}}NVMe\\ Server\end{tabular}} &
        \textbf{\begin{tabular}[c]{@{}c@{}}Data-\\set\end{tabular}} &
          \multicolumn{2}{c}{\textbf{Giraph-OOC}} & \multicolumn{2}{c}{\textbf{TeraHeap}}                                  \\
			\multicolumn{1}{r|}{\textbf{GB}} & \textbf{DRAM} & \textbf{Size}                                               & \textbf{Heap}                          & \textbf{DR2}                          & \textbf{H1} & \textbf{DR2} \\
			\toprule
			PageRank (PR)                    & 85                                                             & 31                                                          & 70                                      & 15                                  & 50              & 35        \\
      \begin{tabular}[c]{@{}l@{}}Community Detection \\ Label Propagation (CDLP) \end{tabular}          & 85                                                             & 31                                                          & 70                                      & 15                                  & 60              & 25        \\
        \begin{tabular}[c]{@{}l@{}}Weakly Connected \\ Components (WCC) \end{tabular}         & 85                                                             & 31                                                          & 70                                      & 15                                  & 60              & 25        \\
			Breadth-first Search (BFS)       & 65                                                             & 31                                                          & 48                                      & 17                                  & 35              & 30        \\
			Shortest Path (SSSP)             & 90                                                             & 31                                                          & 75                                      & 15                                  & 50              & 40        \\
			\bottomrule
		\end{tabular}%
	}
  \caption{Giraph-OOC and \name{} configurations for each workload in
  NVMe server.}
  \label{tab:giraph_conf}
\end{table}

In Spark-SD, to capture the effect of large datasets and limited DRAM
capacity~\cite{chen:hybrid}, we use a small heap size that caches a
limited number of RDDs on-heap and the rest off-heap (fourth column
in Table~\ref{tab:spark_conf}). In Spark-MO we find and use the minimum heap
size that fits all the cached data on-heap (fifth column in
Table~\ref{tab:spark_conf}). In Giraph-OOC, we experimentally find the
minimum heap size for each workload (fourth column in
Table~\ref{tab:giraph_conf}). \name{} uses the same amount of DRAM as
Spark-SD and Giraph-OOC but divides the capacity into H1 and DR2 (used
for Spark and Giraph drivers and I/O page
cache). For the division of DRAM, we explore H1 sizes between 50\%
and 90\% of DRAM capacity, and we report results with a configuration
hand-tuned for each workload. We omit exploration results due to space
constraints. The sixth column in Table~\ref{tab:spark_conf} and
Table~\ref{tab:giraph_conf} show the H1 size of \name{} in each
workload, in Spark and Giraph, respectively.  DR2 is always 16\:GB for
Spark, whereas the fifth and seventh column in Table~\ref{tab:giraph_conf}
show the DR2 size for Giraph.
We limit the available
DRAM capacity in our experiments using \emph{cgroups}. The second
column in Table~\ref{tab:spark_conf} and Table~\ref{tab:giraph_conf}
shows total DRAM capacity in the NVMe server for each workload.

\paragraph{Workloads and datasets:}
We use ten memory-intensive workloads from the Spark-Bench
suite~\cite{Li:SparkBench} and five workloads from the LDBC
Graphalytics suite~\cite{iosup:ldbc} for Giraph. We synthesize
datasets for Spark workloads with the SparkBench data generators. For
Giraph workloads, we use the datagen-9\_0-fb
dataset~\cite{LDBCGrap28:online}. The first and the third columns in
Table~\ref{tab:spark_conf} and Table~\ref{tab:giraph_conf} depict the
workloads and the dataset size.

\paragraph{Execution time breakdowns and \ser overhead:}
We repeat each experiment five times and report the average end-to-end
execution time. We break execution time into four components:
\emph{other} time, \ser + I/O time, minor \gc time, and major \gc
time. \emph{Other} time includes mutator threads time. In \name{}, the
\emph{other} time potentially includes I/O wait due to page faults
when accessing the H2 backing device. In Spark-SD (see
Table~\ref{tab:baselines}), \ser time includes \ser time both for
shuffle and caching. In \name{} and Spark-MO (see
Table~\ref{tab:baselines}), all \ser time is due to shuffling. The JVM
reports the time spent in minor and major \gc.  

To estimate the \ser overhead, which occurs in mutator threads, we use
a sampling profiler~\cite{async-profiler} to collect execution samples
from the mutator threads. The samples include the stack trace, similar
to the flame graph~\cite{flamegraphs} approach. Then we group the
samples for all the paths that originate from the top-level
\texttt{writeObject()} and \texttt{readObject()} methods of the
KryoSerializationStream and KryoDeserializationStream classes. These
samples include both \ser for the compute cache and the shuffle
network path of Spark. We then use the ratio of \ser samples to the
total application thread samples as an estimate of the time spent in
\ser, and we plot this separately in our execution time breakdowns. We
run the profiler with a 10\:ms sampling interval, verifying that this
does not create significant overhead (less than 2\% of total execution
time). 

\section{Evaluation}
\label{sec:eval}

\subsection{Performance Under Fixed DRAM Size}
\label{sec:dram}

\begin{figure*}[t]
	\captionsetup[subfigure]{labelformat=empty}
	\centering
	\includegraphics[width=0.5\linewidth]{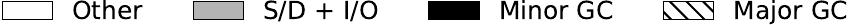}
	\\
	\vspace{-11pt}
	\subfloat[Spark-PR]{\includegraphics[width=0.19\linewidth]{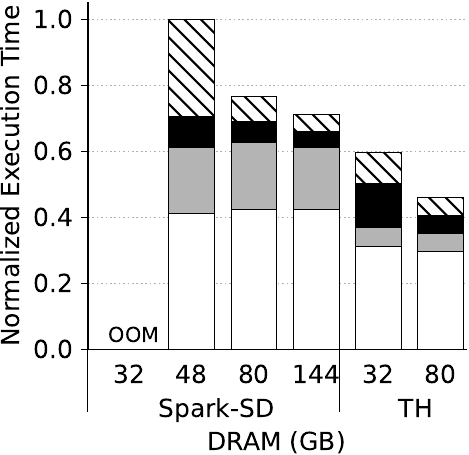}}
	\hspace{2pt}
	\subfloat[Spark-CC]{\includegraphics[width=0.19\linewidth]{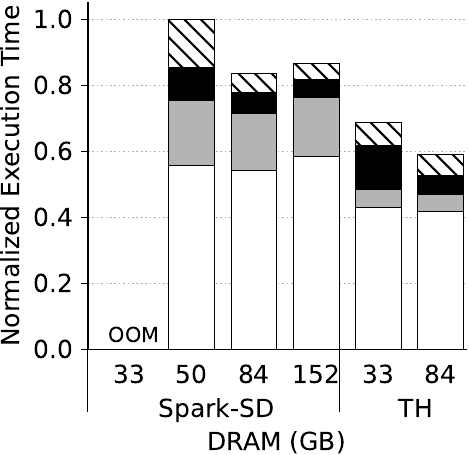}}
	\hspace{2pt}
	\subfloat[Spark-SSSP]{\includegraphics[width=0.19\linewidth]{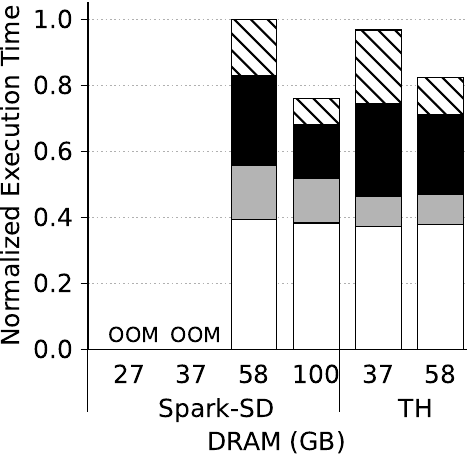}}
	\hspace{2pt}
	\subfloat[Spark-SVD]{\includegraphics[width=0.19\linewidth]{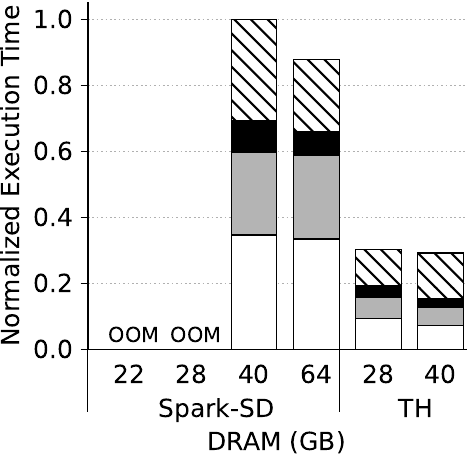}}
	\hspace{2pt}
	\subfloat[Spark-TR]{\includegraphics[width=0.19\linewidth]{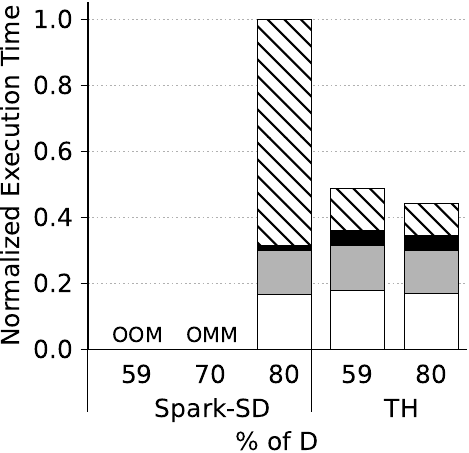}}
	\\
	\vspace{-10pt}
  \subfloat[Spark-LR]{\includegraphics[width=0.19\linewidth]{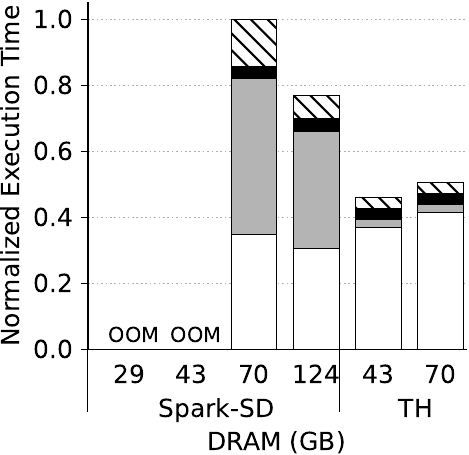}}
	\hspace{2pt}
  \subfloat[Spark-LgR]{\includegraphics[width=0.19\linewidth]{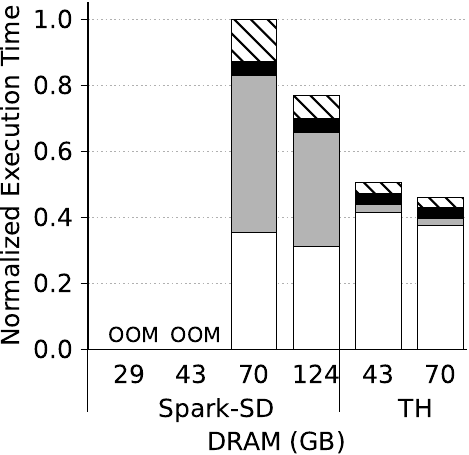}}
	\hspace{2pt}
  \subfloat[Spark-SVM]{\includegraphics[width=0.19\linewidth]{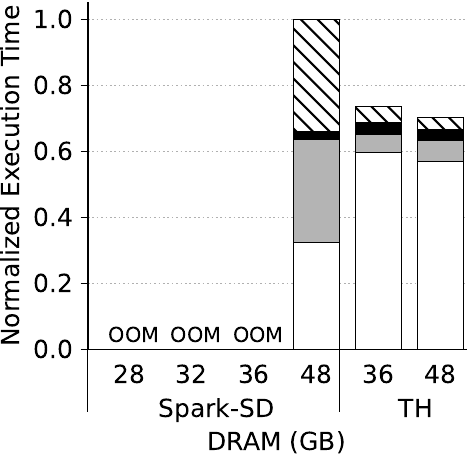}}
	\hspace{2pt}
	\subfloat[Spark-BC]{\includegraphics[width=0.19\linewidth]{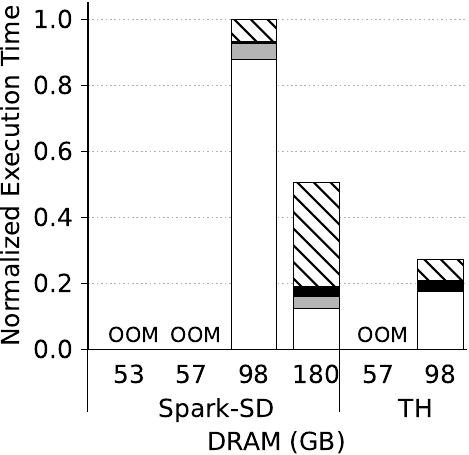}}
	\hspace{2pt}
	\subfloat[Spark-RL]{\includegraphics[width=0.19\linewidth]{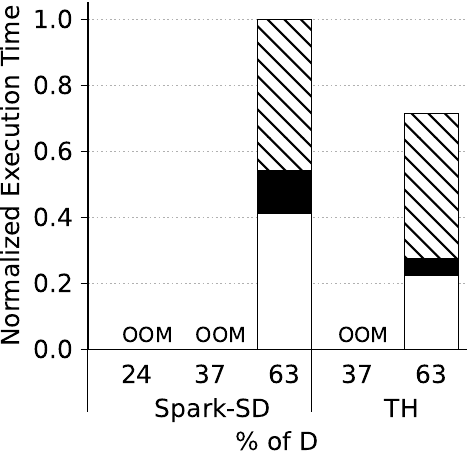}}
	\\
	\vspace{-12pt}
	\subfloat[Giraph-PR]{\includegraphics[width=0.19\linewidth]{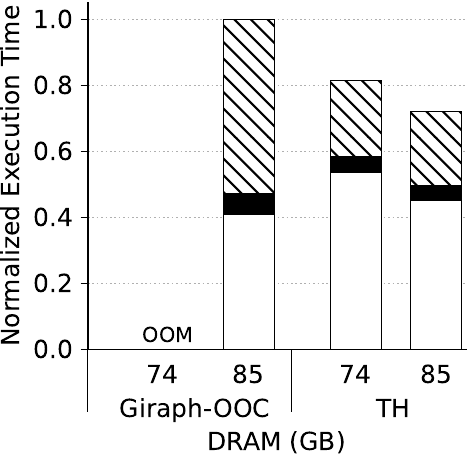}}
	\hspace{2pt}
	\subfloat[Giraph-CDLP]{\includegraphics[width=0.19\linewidth]{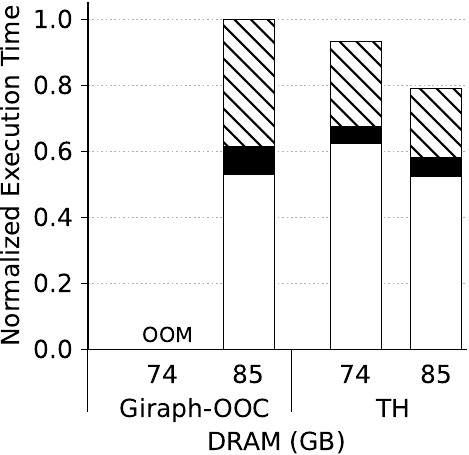}}
	\hspace{2pt}
	\subfloat[Giraph-WCC]{\includegraphics[width=0.19\linewidth]{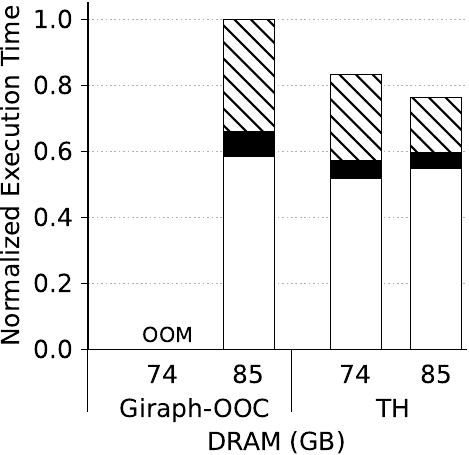}}
	\hspace{2pt}
	\subfloat[Giraph-BFS]{\includegraphics[width=0.19\linewidth]{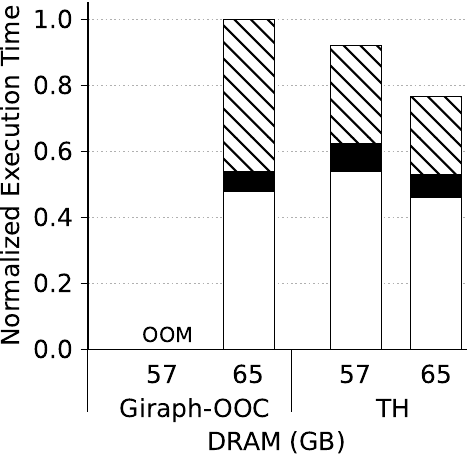}}
	\hspace{2pt}
	\subfloat[Giraph-SSSP]{\includegraphics[width=0.19\linewidth]{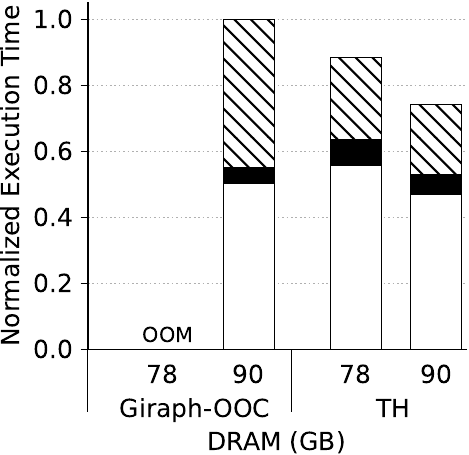}}

  \caption{Overall performance of \name{} (TH) compared to Spark-SD
  and Giraph-OOC in the NVMe server.}
	\label{fig:dram}
\end{figure*}

First, we investigate the performance benefits of \name{} under a
fixed DRAM size. Figure~\ref{fig:dram} shows the performance of
\name{} compared to Spark-SD and Giraph-OOC for the NVMe SSD setup. We
normalize the execution time to the first bar in each figure. Missing
bars indicate out-of-memory (OOM) errors.

Using the same DRAM size, \name{} reduces execution
time in Spark between 18\% (SSSP) and 73\% (BC) compared to Spark-SD.
In Giraph, \name{} reduces execution time between 21\% (CDLP) and 28\%
(PR). In both cases, the performance improvement results from reducing
the \gc overhead, by up to 96\% and 54\% in Spark and Giraph,
respectively. This overhead occurs mainly because cached objects in Spark
and messages and edges in Giraph occupy almost half of the heap,
triggering \gc more frequently. \name{} transfers objects to H2,
stressing H1 less.

In addition, \name{} reduces \ser cost in Spark-SD, between 2\% (BC)
and 93\% (LR), as it provides direct access to deserialized objects in
H2. Note that \ser cost in TR and BC for \name{} is similar to
Spark-SD because the cached data fits in the on-heap cache. In Giraph, the
impact of \name{} on \ser overhead (part of \emph{other}) is minimal
because Giraph serializes objects in the managed heap as well, and not
only as part of moving objects off-heap. Also, in LR, LgR, and SVM
\emph{other} time with \name{} increases by up to 43\% compared to
Spark-SD. These workloads perform streaming access on cached RDDs
elements in each iteration of the ML training phase, which is the
largest part of the execution (100 iterations). Thus, they do not
exhibit locality in the I/O page cache, fetching data from the storage
device during the computation. The average read throughput in these
workloads is 2.9\:GB/s, which is the peak device read throughput.
Using more NVMe SSDs can reduce \emph{other} time for LR, LgR, and
SVM.

\begin{figure}[t]
	\centering
	\vspace{-8pt}
	\subfloat[Spark-SD]{\includegraphics[width=0.49\linewidth]{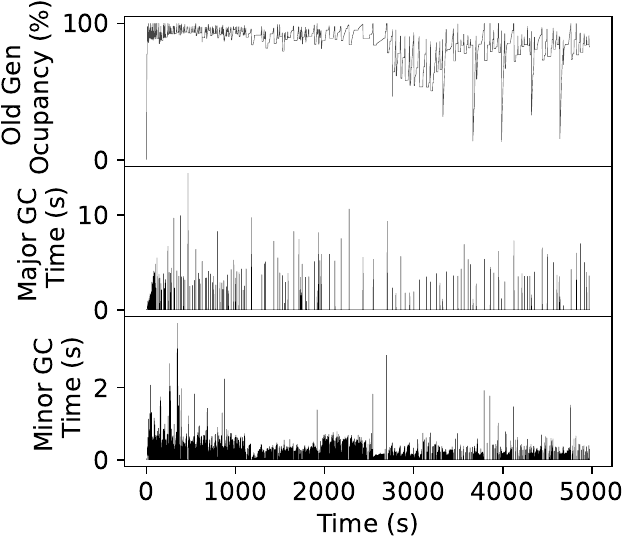}}
	\hspace{1pt}
	\subfloat[\name{}]{\includegraphics[width=0.49\linewidth]{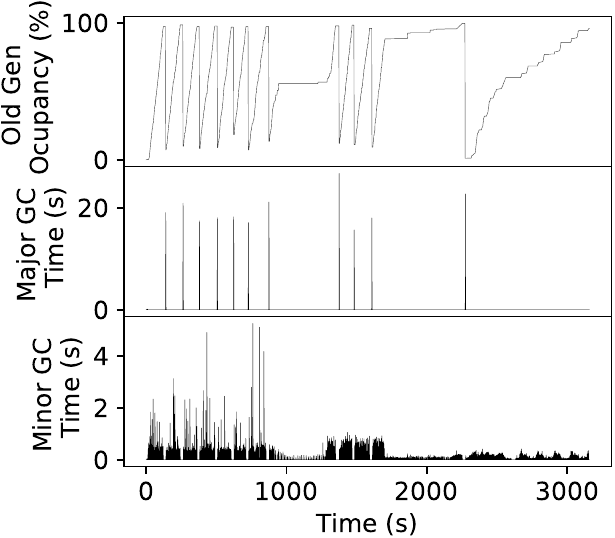}}
	\vspace{-0.2\baselineskip}
	\caption{\gc time and old generation occupancy in PR for (a) Spark-SD
		and (b) \name{}. Heap size = 64\:GB.}
	\label{fig:jstat}
\end{figure}

To examine pressure on the managed heap, Figure~\ref{fig:jstat} shows
\gc behavior for PR with Spark-SD and \name{} with a 64\:GB heap. We
examine the execution time for each minor and major \gc cycle and
monitor the percentage of the old generation consumed by cached
objects. We note that Spark-SD suffers from frequent major \gc cycles.
There are 171 major \gc cycles, each requiring, on average, 3.7\:s.
Each cycle in Spark-SD reclaims 10\% of the old generation objects
(0-3000\:s in Figure~\ref{fig:jstat}), as the remaining objects are
live cached objects. However, \name{} performs only 13 major \gc
cycles. Each cycle in \name{} takes, on average, 16\:s. More than 70\%
is due to I/O during the compaction phase of major \gc. Finally,
moving objects directly from the young generation to H2 reduces total
minor \gc time by 38\% compared to Spark-SD. This reduction is because
\name{} scans fewer cards that track old-to-young references than
Spark-SD. We omit similar results for other workloads due to space
constraints.

\paragraph{Reducing DRAM capacity demands:}
We examine the potential benefit of \name{} in reducing DRAM capacity
demands in Figure~\ref{fig:dram}. Using between $2\times$ and
$8\times$ less DRAM, \name{} outperforms by up to 65\% (SVD) compared
to Spark-SD. In Giraph, \name{} with $1.2\times$ less DRAM improves
performance between 7\% (CDLP) and 18\% (PR). For example, using
\name{} in Giraph-PR, the heap usage in the first phase of the
application (0-330\:s) is between 70\% and 100\%. Then, at the end of
the fifth major \gc, \name{} reduces heap usage to 13\% because it
moves 17\:GB of objects to H2. By reducing memory pressure in H1,
\name{} with less DRAM can provide similar or higher performance than
Spark-SD and Giraph-OOC.

\begin{figure}[t]
	\centering
	\includegraphics[width=\linewidth]{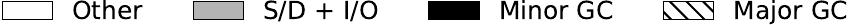}
	\\
	\includegraphics[width=\linewidth]{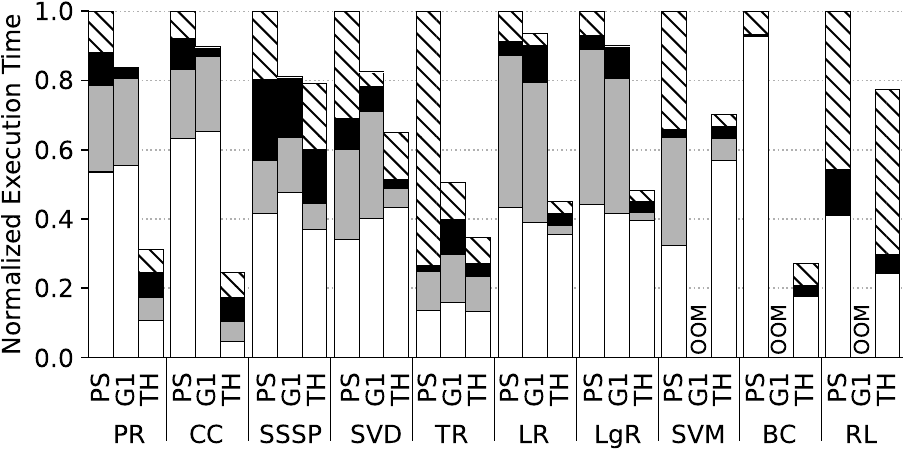}
  \caption{Performance of \name{} (TH) compared to Parallel Scavenge
  (PS) on OpenJDK11 and Garbage First (G1) on OpenJDK17 garbage
  collectors over NVMe server.}
	\label{fig:g1}
\end{figure}

\paragraph{Comparison with newer garbage collectors:}

We next present the performance of \name{} compared to an optimized
version of PS on OpenJDK11 and G1 on OpenJDK17. G1 is a generational,
region-based garbage collector which uses concurrent and parallel
phases to reduce pause time and to maintain high \gc throughput. When
G1 determines that a \gc is necessary, it collects the regions with
the least live data first (garbage first).  Figure~\ref{fig:g1} shows
the performance of Spark with PS, G1, and \name{}, for the same amount
of DRAM.

G1 outperforms PS between 7\% (LR) and 72\% (TC) because it reduces
\gc time by up to 95\%. However, G1 cannot eliminate the high \ser (up
to 44\%) caused by the limited DRAM size and the amount of cached
data. Unlike G1, \name{} eliminates \ser overhead, providing direct
access to the storage resident objects. Thus, \name{} improves
performance between 21\% (CC) and 48\% (LgR) compared to G1.

Note that G1 cannot run SVM, BC, and RL due to fragmentation problems
caused by \emph{humongous objects}. Humongous objects in G1 are these
that are bigger than half of the G1 region size. Such objects are
allocated separately in contiguous regions (humongous regions). A
humongous region can accommodate only one humongous object. The space
between the end of the humongous object and the end of the humongous
region, which in the worst case can be close to half the region size,
is unused. Therefore, when many long-lived humongous objects exist, G1
exhibits significant fragmentation, resulting in OOM errors. PS
resolves fragmentation, performing object compaction when the heap
becomes full. 

We note that \name{} can also be used with G1 to eliminate \ser cost
and reduce the amount of data subject to \gc, by moving long-lived,
humongous objects to H2.
 
\begin{figure}[t]
	\centering
	\includegraphics[width=0.7\linewidth]{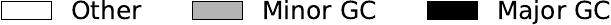}
	\\
	\vspace{-12pt}
	\subfloat[Effect of transfer hint]{\includegraphics[width=0.65\linewidth]{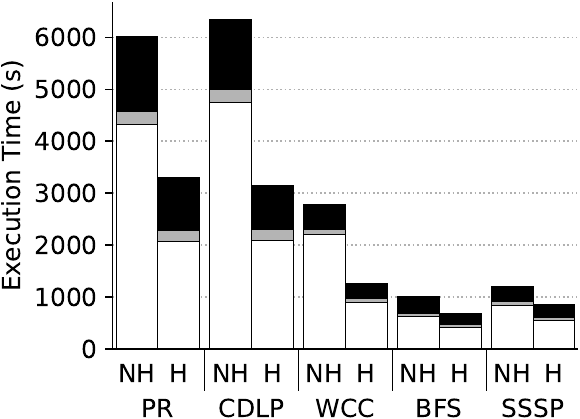}}
	\subfloat[Effect of low threshold]{\includegraphics[width=0.35\linewidth]{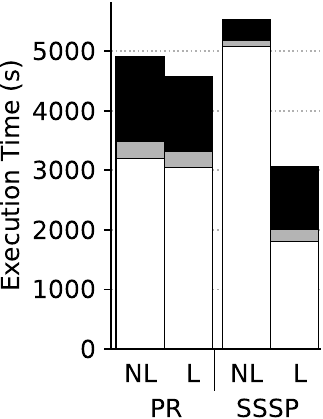}}
	\caption{\name{} performance for Giraph (a) with (H) and without
		(NH) transfer hint, and (b) with (L) and without (NL) low transfer threshold.}
	\label{fig:api}
\end{figure}

\subsection{Effects of Transfer Hint and Low Threshold}
\label{sec:hints}

This section examines the performance effect of using the transfer
hint \texttt{h2\_move()}. Figure~\ref{fig:api}(a) shows the
performance of \name{} with and without using \texttt{h2\_move()}.
Frameworks use \texttt{h2\_move()} to advise \name{} when to move
objects with a specified label to H2. Note that in case of high memory pressure,
\name{} moves to H2 all marked objects without waiting for
\texttt{h2\_move()} hints. Given that \name{} can use only the high
threshold mechanism to decide when to move objects to H2, we explore
eliminating \texttt{h2\_move()}. This results in objects staying
longer in H1. With a high threshold of 85\%, we see
(Figure~\ref{fig:api}(a)) that using \texttt{h2\_move()} improves
\name{} performance between 29\% (SSSP) and 55\% (WCC) compared to not
using the hint. In WCC, using \texttt{h2\_move()}, we move objects to
H2 on average every 215\:s, reducing \gc cost by 39\% compared to not
using the transfer hint, which transfers objects on average every
485\:s. Moving objects with frequent updates to H2 increases
\emph{other} time by up to 59\% (WCC) due to the large cost of
read-modify-write operations on an I/O device. This increases device
traffic by up to 98\% (writes) due to page-based accesses to the
device. Thus, using the transfer hint is necessary to delay
moving objects with frequent updates to H2 until they become
immutable.

Next, we study how effective is the \name{} low threshold mechanism.
Figure~\ref{fig:api}(b) shows \name{} performance using
\texttt{h2\_move()} with and without a low threshold. We use a low
threshold of 50\% (and we leave the high threshold to 85\%). \name{}
will move objects until it reduces H1 usage to 50\%. We use PR and
SSSP with a large dataset (91\:GB) in Giraph. These two workloads
trigger the high threshold mechanism. We use 170\:GB DRAM and 200\:GB
DRAM in PR and SSSP, respectively. The percentage of DR1 over total
DRAM is similar to the corresponding workload in
Figure~\ref{fig:api}(a).

Using a low transfer threshold improves \name{} performance by up to
44\% (SSSP), compared to using only the transfer hint with the high
threshold. For example, in SSSP, during graph loading, we detect high
memory pressure in the fourth major \gc. After the fourth major \gc,
most objects in H1 are related to marked edges. Then, in the fifth
major \gc, we move 44\:GB of marked objects to H2, reducing H1 usage to
50\%. Therefore, the low transfer threshold reduces read-modify-write
operations on the device by up to 95\%, decreasing the \emph{other}
time by up
to 65\%. Although there may be benefits in setting the low and high
thresholds dynamically, we leave this for future work.

\subsection{Storage Capacity Consumption}
\label{sec:overhead}

\begin{table}[]
  \centering
  \resizebox{\columnwidth}{!}{%
    \begin{tabular}{l|ccccccccc}
      \multicolumn{1}{l|}{\textbf{Region Size (MB)}}  &   1 &   2 &   4 &  8 & 16 & 32 & 64 & 128 & 256 \\ \hline
      \textbf{MetaData Size (MB)}& 417 & 209 & 104 & 52 & 26 & 13 &  7 &   3 &   2 \\ 
    \end{tabular}%
    }
    \caption{Metadata size per TB of H2 regarding region size.} 
    \label{tab:dram}
\end{table}

\begin{figure}[t]
	\centering
  \includegraphics[width=0.85\linewidth]{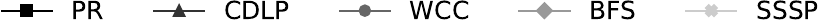}
  \vspace{-8pt}
  \\
  \subfloat[Region size = 16\:MB]{\includegraphics[width=0.48\linewidth]{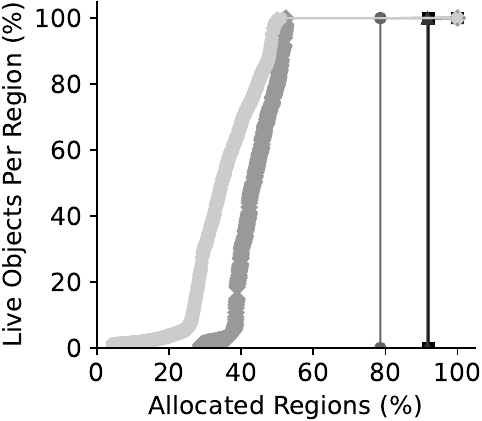}}
  \hspace{3pt}
  \subfloat[Region size = 256\:MB]{\includegraphics[width=0.48\linewidth]{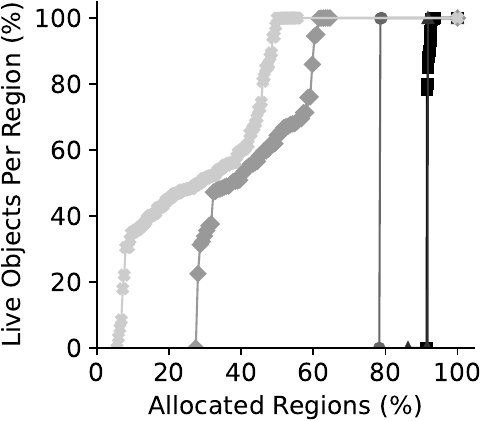}}
  \\
  \vspace{-8pt}
  \subfloat[Region size = 16\:MB]{\includegraphics[width=0.48\linewidth]{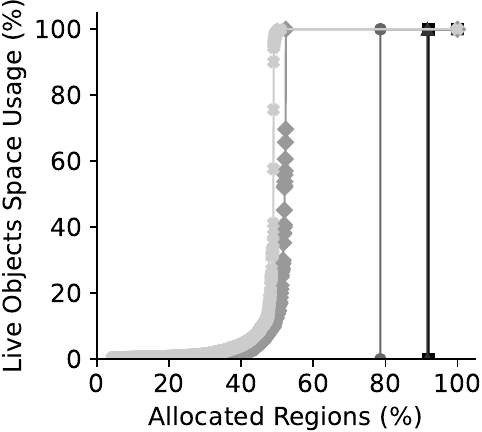}}
  \hspace{3pt}
  \subfloat[Region size = 256\:MB]{\includegraphics[width=0.48\linewidth]{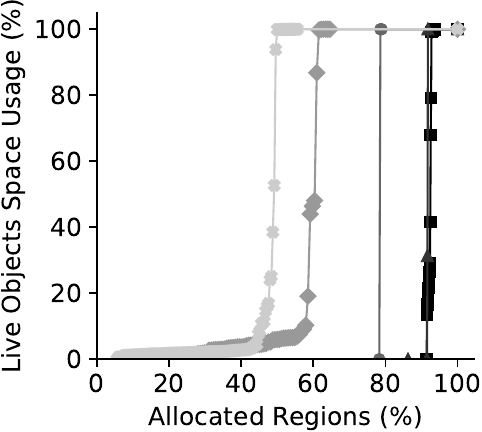}}
  \caption{
    CDF of the percentage of live objects (Figures (a,b)) and space
    occupied by live objects (Figures (c, d)) during execution.}
    \label{fig:alloc_stats}
\end{figure}

This section investigates the storage requirements of H2.
\name{} organizes object groups with a similar lifetime in fixed-size
regions in H2 and reclaims them in bulk. This approach may result in a
waste of space for two reasons. First, when the number of objects in a
group is small, unused space in the corresponding region can be
large. Second, one live object can keep the entire region alive,
preventing \name{} from reclaiming it. Generally, a smaller
region size reduces both of these factors at the cost of
increasing metadata in DRAM.

We first show how the region size affects the metadata size for H2.
Table~\ref{tab:dram} shows the metadata size in DRAM per TB of H2, for
region sizes between 1\:MB and 256\:MB. As we increase the region size
from 1\:MB to 256\:MB, the total metadata in DRAM decreases from
417\:MB to 2\:MB.

Figures~\ref{fig:alloc_stats}(a) and~\ref{fig:alloc_stats}(b) show the
CDF of the percentage of live objects per region for \emph{all
allocated} regions with 16\:MB and 256\:MB size, respectively.
Figures~\ref{fig:alloc_stats}(c) and~\ref{fig:alloc_stats}(d) show the
CDF of the percentage of space occupied by live objects for 16\:MB and
256\:MB regions. The number of allocated regions is equal to the sum
of reclaimed regions during execution and the active regions before
JVM shutdown. Although not shown in these figures, we observe in our
measurements that unused space is between 1\% and 3\% for all
workloads in both region sizes. Essentially, \name{} is able to use
the space in each region it allocates with its append-only placement.

In Figures~\ref{fig:alloc_stats} (a) and~\ref{fig:alloc_stats}(b) all
regions with 0\% live objects are reclaimed during execution. We see
that in PR, CDLP, and WCC, \name{} reclaims most allocated regions and
around 90\% in CDLP and PR for both region sizes.  In BFS and SSSP,
\name{} reclaims 28\% and 6\% of the total allocated regions,
respectively. In BFS and SSSP, although most of the objects in a
region are live, most of the space is occupied by large dead arrays.
For example, in SSSP with 256\:MB regions, in 90\% of the regions at
least 20\% of the objects are live. However, in 45\% of the regions,
the live objects occupy less than 10\% of the allocated region space
(Figure~\ref{fig:alloc_stats}(d)). In these cases, using 16\:MB
regions is more appropriate because they reduce by 10\% (BFS) the
space waste compared to 256\:MB regions. We believe that future work
can investigate object placement policies for H2 that takes into
account object size to further improve space efficiency on storage
devices.

\begin{figure}[t]
	\centering
	\subfloat[]{\includegraphics[width=0.49\linewidth]{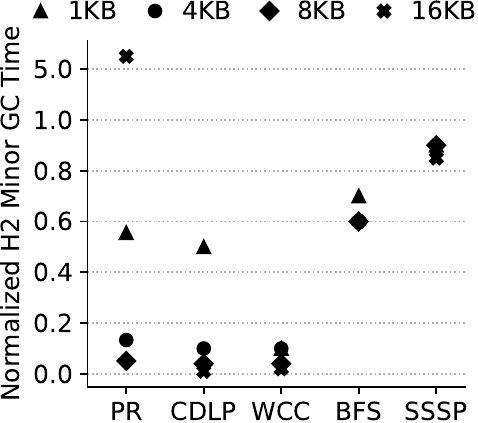}}
	\subfloat[]{\includegraphics[width=0.49\linewidth]{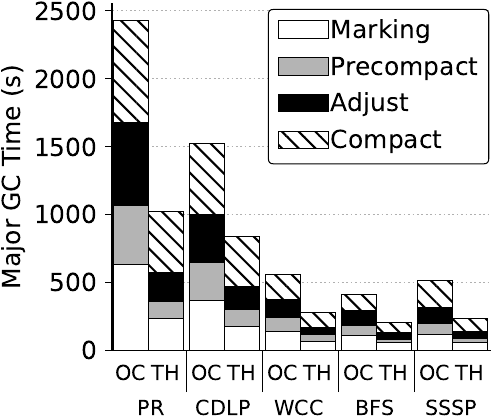}}
	\caption{(a) Minor \gc time in H2 for different card segment sizes
		using a 256\:MB stripe size in Giraph. (b) Major \gc time using
		Giraph-OOC (OC) and \name{} (TH).}
	\label{fig:gc}
\end{figure}

\begin{figure*}[t]
	\centering
	\includegraphics[width=0.5\linewidth]{fig/fig9legend-crop.pdf}
	\\
	\vspace{-12pt}
	\subfloat[Spark-SD vs \name{}]{\includegraphics[width=0.33\linewidth]{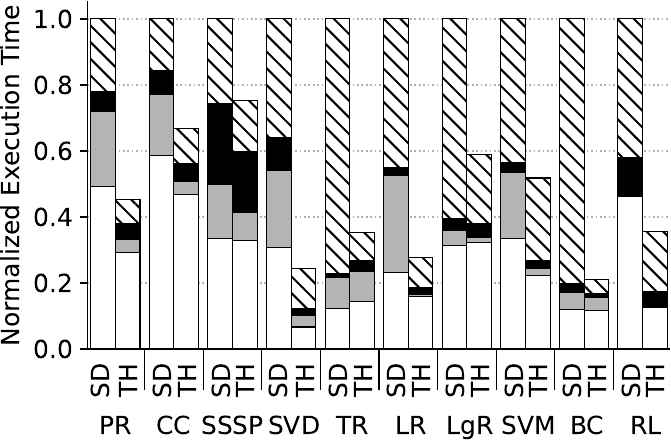}}
  \hspace{0.2pt}
	\subfloat[Spark-MO vs \name{}]{\includegraphics[width=0.33\linewidth]{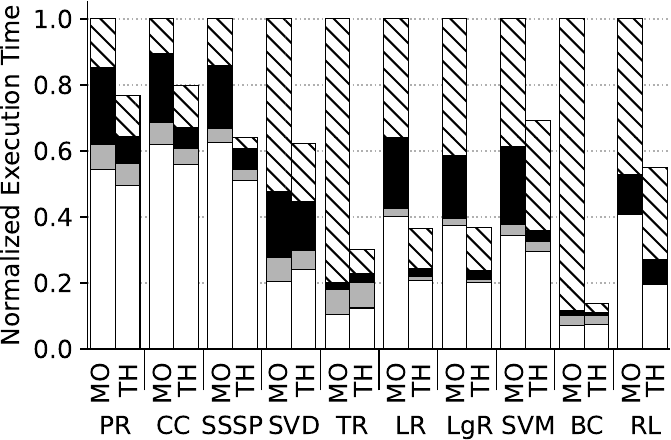}}
  \hspace{0.2pt}
  \subfloat[Panthera vs \name{}]{\includegraphics[width=0.33\linewidth]{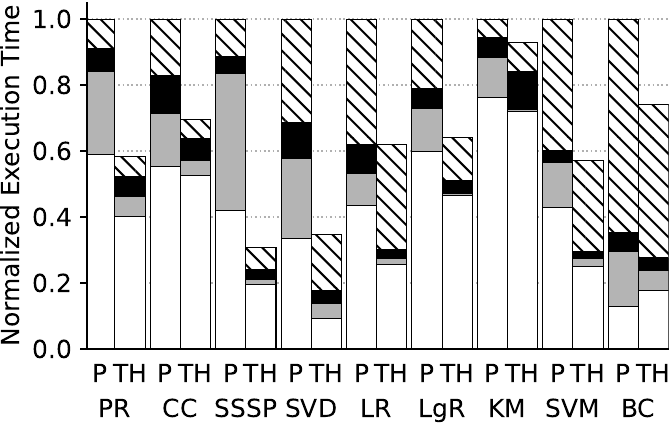}}
	\caption{\name{} performance (TH) compared to (a) Spark-SD,
		(b) Spark-MO, and (c) Panthera (P) over NVM server.}
	\label{fig:nvm}
\end{figure*}

\subsection{\gc Overhead}
The garbage collector in \name{} performs additional work during minor
\gc that involves scanning H2 cards and updating backward references.
We evaluate this overhead for different card segment sizes in Giraph.
Figure~\ref{fig:gc}(a) shows minor \gc time in H2 for 1\:KB, 4\:KB,
8\:KB, and 16\:KB card segments, normalized to 512\:B card segments.
We see that increasing the size of card segments from 512\:B to 16\:KB
reduces minor \gc time on average by $64\%$. Larger card segments
result in a smaller card table, and less time is required to scan the
respective cards. However, increasing card segment size increases the
cost of scanning each card segment if the respective card is marked as
dirty. For example, increasing the card segment size in PR  from
512\:KB to 16\:KB leads to an increase in minor \gc time for scanning
and update H2 objects with backward references (H2 to H1) by
$5\times$. In Spark, updates to H2 objects are infrequent compared to
Giraph, as RDDs are immutable.

Next, we examine the overheads introduced by \name{} during major \gc
for H1 by moving objects to H2, which involves device I/O when using
SSDs as the backing device. Figure~\ref{fig:gc}(b) shows the four
phases of major \gc time using Giraph-OOC and \name{}. Overall,
\name{} improves all phases of major \gc by up to 75\% (BFS) compared
to Giraph-OOC because we avoid scanning H2 objects. For example, in
PR, the collector avoids following in each GC, on average, 109 million
forward references from H1 to H2 objects. We note that the compaction
phase takes between 37\% and 44\% of the major \gc time in \name{} due
to the device I/O.

\subsection{\name{} Performance with NVM}
\label{sec:nvm}

Figure~\ref{fig:nvm} shows the performance of Spark-SD, Spark-MO, and
\name{} in our NVM-based setup. We present only Spark workloads due to
space constraints. Our goal is to examine the benefits of \name{} when
using NVM to increase the heap size, which can eliminate \ser at
increased \gc cost for native. Figure~\ref{fig:nvm}(a) shows that
\name{} improves performance by up to 79\% and on average by 56\%,
compared to Spark-SD. Unlike the off-heap cache in Spark-SD, \name{}
allows Spark to directly access cached objects in H2 via load/store
operations to NVM, without the need to perform \ser. \name{}
significantly reduces \ser and \gc time compared to Spark-SD by up to
97\% and 93\%, respectively. 

Figure~\ref{fig:nvm}(b) shows that \name{} improves performance by up
to 86\% and on average by 48\%, compared to Spark-MO. The main
improvement of \name{} results from the reduction of minor \gc and
major \gc time by up to 88\% (on average by 52\%) and 96\% (on average
by 46\%) compared to Spark-MO, respectively. In Spark-MO, running the
garbage collector on top of NVM (using DRAM as a cache) incurs high
overhead due to the latency of NVM~\cite{izra:nvm} and the agnostic
placement of objects. For instance, minor \gc time in Spark-MO
increases on average by 36\% compared to Spark-SD
(Figure~\ref{fig:nvm}b) because objects of the young generation are
placed in NVM, resulting in higher access latency for the garbage
collector. Unlike \name{} that controls object placement in NVM (H2),
Spark-MO relies on the memory controller to move objects between DRAM
and NVM. We measure that Spark-MO incurs on average $5.3\times$ and
$11.8\times$ more read and write operations to NVM compared to
\name{}, resulting in higher overhead. Thus, the ability to maintain
separate heaps allows \name{} to both limit \gc cost and reduce the
adverse impact of the increased NVM access latency on \gc time.

We also compare \name{} with Panthera~\cite{wang:Panthera}\footnote{As
Panthera is not publicly available, we are thankful to the authors for
providing us their code.}, a system designed to use NVM as a heap in
Spark. Panthera extends the managed heap over DRAM and NVM, placing
the young generation in DRAM and splitting the old generation into
DRAM and NVM components. We configure Panthera as Wang \emph{et.
al}~\cite{wang:Panthera} report: We use a 64\:GB heap, 25\% on DRAM
(16\:G), and 75\% on NVM. We set the size of the young generation to
$\frac{1}{6}$ (10\:GB) of the total heap size and place it entirely on
DRAM. We set the size of the old generation to the rest of the heap
size (54\:GB) and place 6\:GB on DRAM and the rest (48\:GB) on NVM. We
configure \name{} to use an H1 of 16\:GB and map H2 to NVM. Thus, both
systems use the same DRAM and NVM capacity.

Figure~\ref{fig:nvm}(c) shows that \name{} improves performance
between 7\% and 69\% compared to Panthera across all workloads.
Panthera bypasses the allocation of some objects in the young
generation, allocating them directly to the old generation. However,
each major \gc still scans all objects in the old generation, which
increases overhead as the heap address space grows. Instead, \name{}
reduces the address space that needs to be scanned by the garbage
collector. Note that Panthera increases the accesses to NVM because it
allocates mature long-lived objects that are highly read and updated
by the mutator threads. Specifically, it increases \emph{other} by up
to 53\% because it performs more NVM read (up to $54\times$) and NVM
write (up to $51\times$) operations, than \name{}.

\subsection{Performance Scaling}
\label{sec:scaling}

\begin{figure}[t]
	\centering
	\vspace{-5pt}
	\subfloat[Number of mutator
		threads]{\includegraphics[width=0.49\linewidth]{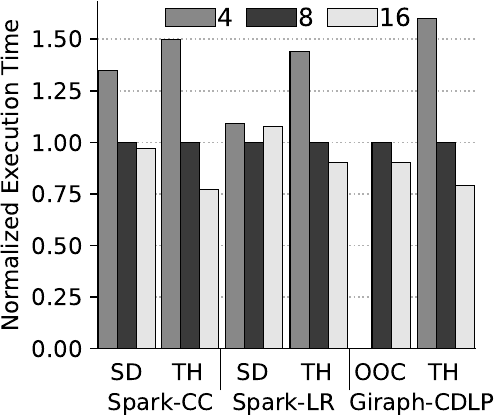}}
	\subfloat[Dataset size (GB)]{\includegraphics[width=0.49\linewidth]{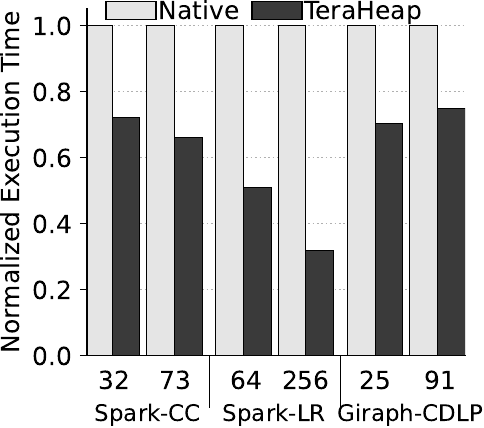}}
  \caption{Performance scaling with (a) number of mutator threads and
  (b) dataset size in the NVMe server.}
	\label{fig:scala}
\end{figure}

A benefit of \name{} is that it allows increasing the number of
mutator threads in Spark and Giraph executors. In both Spark and
Giraph, each mutator thread processes a separate partition. Thus, as
the number of threads in the executor increases, the object allocation
rate increases, leading to higher \gc{} cost.
Figure~\ref{fig:scala}(a) shows the performance of CC, LR, and CDLP
(other workloads show similar behavior) using Spark-SD, Giraph-OOC,
and \name{} (TH) with 4, 8, and 16 threads, normalized to 8 threads
per configuration. We note that Giraph-OOC with four threads results
in an OOM error. \name{} allows applications to scale performance
further to 23\% with $2\times$ more threads. However, Spark-SD does
not scale beyond 8 threads in LR because \gc{} cost increases (by
44\%), eliminating any benefits from using more threads. We note that
increasing the number of threads in Spark-SD reduces \ser cost by up
to 55\% (CC) because Spark parallelizes the \ser process. Although
Giraph-OOC (native) improves performance by 10\% using 16 executor
threads, it still performs $1.4\times$ more major GCs than eight
executor threads. Finally, \name{} significantly alleviates memory
pressure by moving a large portion of H1 objects to H2, leaving more
room for mutator threads to work without the need for frequent \gc.

We also investigate the performance benefits of \name{} for a larger
dataset in Figure~\ref{fig:scala}(b). We observe similar (CDLP) or
higher improvements (CC, LR) compared to the smaller datasets. \name{}
is robust to different dataset sizes and improves performance by up to
70\% compared to Spark-SD and Giraph-OOC, while our expectation is
that benefit will increase further as dataset size increase.

\section{Related Work}
\name{} combines techniques from several areas, including memory
management and storage. Thus, we group the related work in the
following categories:                                           
(1) region-based memory management,
(2) scaling managed heaps beyond DRAM, and
(3) mitigating \ser overhead. 

\paragraph{Region-based memory management:}
Managed big data frameworks have started to use region-based memory
management for large heaps. Facade~\cite{nguyen:facade} provides a
compilation framework that transforms programmer-specified classes for
off-heap allocation. However, it increases the programmer's effort
because they need to specify when to free objects from native memory.
Broom~\cite{gog:broom} uses region annotations but requires
refactoring of applications' source code. Yak~\cite{khanh:yak}
requires programmers to annotate epochs in applications. Yak allocates
all objects in an epoch on a second region-based heap to reduce \gc
time.  The epoch abstraction is appropriate for the map-reduce
programming pattern. However, it cannot handle objects computed lazily
or accessed from arbitrary program locations. Deca~\cite{lu:deca}
proposes lifetime-based memory management for Spark. However, their
work only applies to Spark and cannot be used for other frameworks.
Unlike prior work, \name{} requires adding hints only in the framework
layer. Then, \name{} dynamically selects all appropriate objects in
the transitive closure of root objects. NG2C~\cite{lifetime-ferreira}
uses runtime profiling to identify long-lived objects. They incur
online profiling overhead. Others use offline allocation site
profiling to manage objects~\cite{pretenuring,Blackburn:2001:PJ}.
Lifetime profiling complements \name{}, further improving efficiency.

\paragraph{Scaling managed heaps beyond DRAM:}
Recent efforts target NVM for storing managed heaps beyond DRAM. Akram
et al.~\cite{akram:ration, akram:crystal} focus on improving NVM write
endurance. Yang et al.~\cite{nvm-gc-eurosys} report high \gc overhead
with NVM-backed volatile heaps and optimize the G1 \gc for Intel
Optane persistent memory.  Panthera~\cite{wang:Panthera} extends the
managed heap over hybrid DRAM and non-volatile memory (NVM) to scale
on-heap caching in Spark. Panthera increases \gc overhead as scanning
and compacting objects on the managed NVM heap costs more than
collecting the DRAM heap. Also, TMO~\cite{weiner:tmo} monitors
application DRAM usage and transparently offloads cold data to NVMe
SSD. Unlike these works, \name{} control which objects to move to the
second heap and eliminates slow \gc traversals over objects on NVM or
NVMe SSD.

\paragraph{Mitigating \ser overhead:}
Several libraries~\cite{thrift,kryo,protocol-buffers} improve the
efficiency of \ser, but they cannot reduce high \gc cost in big data
frameworks. Skyway~\cite{nguyen:skyway} reduces the \ser cost by
directly transferring objects through the network in distributed
managed heaps, but it does not cope with DRAM limitations and \gc
overheads. SSDStream~\cite{ssdstreamer} is a userspace SSD-based
caching system that uses DRAM as a stream buffer for SSD devices.
Although SSDStreamer reduces \ser cost by providing a lightweight
serializer, it cannot reduce \gc cost and the memory pressure in the
managed heap. Recent work~\cite{jang:arch,pourhabibi:optimus} reduces
\ser overheads in analytics frameworks using custom hardware and
modifications to the programming model. Other
works~\cite{matei:champions, taranov:naos, ibanez:zerializer} focus on
reducing \ser cost by reducing the number of object copies across
buffers. This body of work does not mitigate directly \gc overhead.
\name{} is the first work that eliminates both \gc and \ser in big
data analytics frameworks. \name{} works on commodity hardware and
uses load/store instructions to access objects in the second heap
without data serialization.

\section{Conclusions}

Managed big data analytics frameworks demand large managed heaps as
datasets grow. This work proposes and evaluates \name{}, which extends
the JVM to use a transparent, high-capacity heap over a fast storage
device alongside the regular heap, reducing memory pressure. \name{}
reduces \gc overhead and eliminates \ser cost by fencing the collector
from scanning the second heap and providing direct access to objects
on the second heap. We find that \name{} improves the Spark and Giraph
performance by up to 73\% and 28\%, respectively. Overall, our
proposed approach of managing large memory in the JVM as customized,
separate heaps is a promising direction for incorporating huge address
spaces in managed environments and reducing memory pressure without
incurring high \gc overhead.

\bibliographystyle{plain}
\bibliography{paper}

\end{document}